\documentclass[12pt,letterpaper]{article}
\usepackage{graphicx}
\usepackage{amsmath}
\usepackage{amssymb}
\usepackage{hyperref}
\usepackage{tikz}
\usetikzlibrary{arrows,decorations.pathmorphing,decorations.pathreplacing,decorations.markings,shapes.geometric,calc,positioning,chains,matrix,scopes} 

\numberwithin{equation}{section}

\def\bx{\mathbf{x}}
\def\bq{\mathbf{q}}

\newcommand{\cO}{\mathcal{O}}
\newcommand{\bZ}{\mathbb{Z}}
\newcommand{\bC}{\mathbb{C}}

\newcommand{\CP}{\mathbb{CP}}
\newcommand{\cI}{\mathcal{I}}

\DeclareMathOperator{\vol}{vol}
\newcommand{\tgd}{{{\partial}_B{}}}
\newcommand{\tgdb}{{{\bar{\partial}_B}}}

\newcommand{\noedits}[1]{}

\DeclareMathOperator{\Tr}{Tr}

\DeclareMathOperator{\ind}{ind}

\begin{document}

\begin{titlepage}

\begin{flushright}
IPMU 13-0089 \\
RIKEN-MP-71
\end{flushright}
\vskip 2cm

\begin{center}
{\Large \bfseries

Superconformal Indices and M2-Branes \\

}

\vskip 1.2cm

Richard Eager$^\clubsuit$ and
Johannes Schmude$^\spadesuit$,

\bigskip
\bigskip

\begin{tabular}{ll}
$^\clubsuit$  & Kavli Institute for the Physics and Mathematics of the Universe (WPI), \\
& University of Tokyo,  Kashiwa, Chiba 277-8583, Japan\\
$^\spadesuit$ & RIKEN Nishina Center, Saitama 351-0198, Japan 
\end{tabular}

\vskip 1.5cm

\textbf{Abstract}
\end{center}

\medskip
\noindent
We derive the superconformal index of the world-volume theory on
M2-branes probing the cone over an arbitrary Sasaki-Einstein
seven-manifold.  The index is expressed in terms of the cohomology groups
of the cone. We match our supergravity results with known results from gauge theory.  Along the way we derive the spectrum of short Kaluza-Klein multiplets 
on generic Sasaki-Einstein seven-manifolds.
\bigskip
\vfill
\end{titlepage}

\setcounter{tocdepth}{2}
\tableofcontents
\newpage
\section{Introduction}
The superconformal index of a three-dimensional superconformal field theory is the partition function of the theory on $S^1 \times S^2$ with supersymmetric boundary conditions.  Equivalently, the index can be expressed as the trace over all operators in the theory weighted by their fermion number
\begin{equation}
  \label{eq:the_index}
  \cI(t,z_i) = \Tr[(-1)^F t^{\epsilon + j_3} z_{i}^{h_i}].
\end{equation}
Here $\epsilon$ is the operator dimension, $j_3$ is the spin of the operator, $F$ is its fermion number, and $h_i$ label the charges of the operator under global symmetries.  The superconformal index is invariant under exactly marginal deformations and can be computed from an ultraviolet Lagrangian description, provided that the infrared R-symmetry can be identified in the ultraviolet \cite{Festuccia:2011ws}.  The superconformal index was originally defined for four dimensional superconformal field theories \cite{Kinney:2005ej, Romelsberger:2005eg} and was generalized to three dimensional theories in \cite{Bhattacharya:2008zy}.
 
A large class of three dimensional superconformal field theories is realized as the low energy effective theory of multiple M2-branes probing a Calabi-Yau fourfold singularity.  These theories have $\mathcal{N} = 2$ supersymmetry and have a holographic dual description as M-theory on the product of four-dimensional anti-de Sitter space $AdS_4$ and a seven-dimensional Sasaki-Einstein manifold.
The simplest Calabi-Yau fourfold singularity is $\mathbb{C}^4/\mathbb{Z}_k.$  The theory of M2-branes at this singularity is realized as the low energy limit of a quiver Chern-Simons theory with $\mathcal{N} = 8$ supersymmetry when $k = 1,2$ \cite{Aharony:2008ug}. 
The holographic dual theory is M-theory on $AdS_4 \times S^7/\mathbb{Z}_k$ and the supergravity index was computed by summing all contributions from short multiplets \cite{Bhattacharya:2008zy}.
The field theory and supergravity superconformal indices were shown to match in the large $k$ limit \cite{Bhattacharya:2008bja}.  At finite $k,$ monopole operators contribute to the index and their contribution can be computed using localization \cite{Kim:2009wb, Imamura:2011su}.

In this paper we will derive the gravity superconformal index for any theory of the form $AdS_4 \times SE_7.$
Previously the supergravity index was computed for the homogenous Sasaki-Einstein seven-manifolds using known Kaluza-Klein spectra \cite{Cheon:2011th}.  However, to match the field theory index and the supergravity index, several of the Kaluza-Klein modes in \cite{Merlatti:2000ed} had to be dropped.  Since the spectrum has not been well tested, the authors suggested that the Kaluza-Klein spectrum should be revisited.  We find that a careful analysis of the Kaluza-Klein modes agrees with the field theory index \cite{Gang:2011jj, Cheon:2011th,Imamura:2011uj}.
Our general form of the supergravity index succintctly reproduces previous computations of the gravity index \cite{Cheon:2011th}.
We find complete agreement with previous large-$N$ computations of the index \cite{Gang:2011jj, Cheon:2011th,Imamura:2011uj}.  

We construct the Kaluza-Klein multiplets on $AdS_4$ from various tensors defined on the Sasaki-Einstein manifold following the methodology of \cite{Eager:2012hx}.
Our analysis focuses on generic Sasaki-Einstein manifolds.  Much of our analysis builds upon previous work on Kaluza-Klein spectroscopy for coset manifolds 
\cite{Merlatti:2000ed,Ceresole:1999zg,DAuria:1984vy,Fabbri:1999mk}.

Multiplet shortening and the short multiplets contributing to the index can be described using the tangential
Cauchy-Riemann operator  $\tgdb$ and the associated Kohn-Rossi
cohomology groups $H_\tgdb^{p,q}$. In general, the cotangent bundle
over a Sasaki-Einstein manifold $Y$ can be decomposed as
\begin{equation}
  \Omega_Y = \mathbb{C} \eta \oplus \Omega_Y^{1,0} \oplus \Omega_Y^{0,1}.  
\end{equation}
The operator $\tgdb$ is the projection of the exterior derivative on
$\Omega_Y^{0,1}$, the cohomology of this complex is
$H_\tgdb^{p,q}$ \cite{MR0177135, MR604043}.
The Kohn-Rossi cohomology groups are isomorphic to $H^q(X, \wedge^p
\Omega^\prime_X)$ defined on the cone, where $\Omega^\prime_X$ is the
part of the holomorphic cotangent bundle $\Omega_X$ perpendicular to
the dilatation vector field.  Our main result is a formula for the gravity superconformal index in as a trace over linear combinations of
the groups $H^q(X, \wedge^p \Omega^\prime_X)$.

\paragraph{Organization}

The organization of this paper is as follows: in section \ref{summary} we 
calculate the single trace index drawing on the results from the
remaining sections of this paper. Section \ref{cohomology} shows how the trace over
cohomology groups can be evaluated for toric and Fano manifolds. The
Kaluza-Klein analysis of Sasaki-Einstein seven-manifolds used in
section \ref{summary} is performed in in section
\ref{sec:kaluza-klein-analysis}, with technical details refered to the
appendices \ref{conventions} and
\ref{sec:details-of-supergravity-analysis}. Section
\ref{sec:previous-work} describes relations between this paper and previous work on
Kaluza-Klein spectroscopy.

\section{Calculation of the Index}\label{summary}
In this section we list the multiplicity of each short multiplet
appearing in supergravity solutions of the form $AdS_4 \times SE_7$
and their contribution to the superconformal index.
The single trace superconformal index is defined by the following
modification of \eqref{eq:the_index}
\begin{equation}
  \label{eq:single_trace_index}
  \cI_{s.t.}(t,z_i) = \Tr_{s.t.}[(-1)^F t^{\epsilon + j_3} z_{i}^{h_i}].
\end{equation}
Only states with
\begin{equation}
 \{Q, S\} = \epsilon - j_{3} - y = 0
\end{equation}
contribute, where $y$ is the R-charge. Short multiplets
contributing to the index and their multiplicities are listed in table
\ref{tab:short}.
An element $f$ of cohomology has R-charge $\mathcal{L}_{D} f = 2i D f.$
Here $\mathcal{L}_{D}$ denotes the Lie derivative along the dilation vector field and $2D$ is its corresponding eigenvalue.
We normalize each multiplet so that its primary has R-charge $y.$  The R-charge $y$ differs from the R-charge $2D$ of the corresponding cohomology element by a constant shift.

A short multiplet whose primary has quantum numbers $(y+j_3+1,j_3,y)$ contributes $(-1)^{2j_3 +1} t^{y + 2 j_3 + 2}$ to the index.
The hypermultiplets with quantum numbers $(y,0,y)$ contribute $t^{y}$ to the index.
\begin{table}[htdp]
\begin{center}
\begin{tabular}{|l|c|l|c|c|}
\hline
Multiplet & Primary $(\epsilon, j_3, y)$ & Multiplicity & Index & Index \\
\hline
short graviton 		& $(y+2,1,y)$				& $H^0(X,\wedge^3 \Omega'_X)$		& $- t^{y + 4}$ 	& $-t^{2D+2}$ \\
short gravitino 		& $(y+\frac{3}{2},\frac{1}{2},y)$	& $H^0(X, \Omega'_X)$				& $t^{y + 3}$ 	& $t^{2D+2}$ \\
short vector $Z$/betti    & $(y+1,0,y)$				& $H^1(X,\Omega^\prime_X)$ 			& $-t^{y + 2}$	& $-t^{2D+2}$ \\
short vector $A$ 		& $(y+1,0,y)$				& $H^0(X,\wedge^2 \Omega'_X)$		& $- t^{y + 2}$ 	& $-t^{2D}$\\
hyper	 		& $(y,0,y)$				& $H^1(X,\wedge^2 \Omega'_X)$		& $  t^{y}$ 	& $t^{2D}$\\
hyper	 		& $(y,0,y)$				& $H^2(X, \Omega'_X)$		& $  t^{y}$ 	& $t^{2D+2}$\\
hyper 			& $(y,0,y)$				& $H^0(X,\mathcal{O}_X)$			& $t^{y}$ 		& $t^{2D}$\\
\hline
\end{tabular}
\end{center}
\caption{Short multiplets and their contribution to the index.}
\label{tab:short}
\end{table}
Summing the contributions of the short multiplets, we find that the single particle supergravity index is
\begin{equation} \label{aab}
  \begin{aligned}
  1 +  \cI_{s.t.}(t)   =  \sum \Tr  t^{2D} & \bigm|   H^0(X,\mathcal{O}_X) 
    \ominus H^0(X,\wedge^2 \Omega'_X) 
    \oplus H^1(X,\wedge^2 \Omega'_X) \\
    &  \oplus t^2 H^0(X, \Omega'_X) \ominus t^2 H^1(X,\Omega^\prime_X) \oplus t^2 H^2(X, \Omega'_X) \ominus t^2 H^0(X,\wedge^3 \Omega'_X) .
\end{aligned}
\end{equation}
The single particle index is similar to the single-trace index, but it also includes the derivates of the single-trace operators.  These two indices are related by
\footnote{Here we omit the contribution from the identity operator.}
\begin{equation}
\cI_{s.t.}(t) = (1-t^2) \cI_{s.p.}(t).
\end{equation}
We consider only Sasaki-Einstein manifolds.  For Sasaki-Einstein spaces with singularities, there can be 
additional contributions \cite{Gukov:1998kk, Nakayama:2005mf}.
We next explain how the multiplet shortening conditions arise from the supergravity spectrum after reviewing the structure of short and long multiplets.

\subsection{Unitary Representations of the $\mathcal{N} = 2$ Superconformal Algebra}
Supergravity on $AdS_4 \times SE_7$ has $\mathcal{N} = 2$ superconformal symmetry.  We begin by recalling the properties of the three dimensional $\mathcal{N} = 2$ superconformal group $Osp(2|4)$.  Its bosonic subgroup is $Sp(4, \mathbb{R}) \oplus SO(2)_R.$  The first factor $Sp(4, \mathbb{R}) \cong SO(2,3)$ is the isometry group of $AdS_4$ and the second factor, $SO(2)_R$, is the R-symmetry group.

Unitary representations are labeled by their eigenvalues $\epsilon, j_3, y$ under the dilation, angular momentum, and R-symmetry operators.  Unitary representations with spin $j_3 > 0$ satisfy $\epsilon - j_{3} - y \ge  0.$  Representations saturating this bound have null states and the representation shortens.  Hence such representations are called {\it short}.  Representations not saturating this bound are called {\it long}.
When $j_3 = 0$ representations  satisfying $\epsilon = y$ are called
{\it isolated}.  All other representations with $j_3 = 0$ satisfy
$\epsilon \ge y + 1$; those saturating this bound are called {\it short}.

\subsection{The Short Graviton Multiplet}

The graviton multiplet, shown in table \ref{tab:graviton_multiplet},
can be constructed starting from a scalar eigenfunction of the Laplacian on the
Sasaki-Einstein manifold.  The scalar eigenfunction is the wave function of the spin 2 graviton in the multiplet.    
An eigenvalue $\Delta_0$ of the scalar Laplacian is bounded by its
charge\footnote{$q$ is the eigenvalue of the Lie derivative along $\xi$,
  $\pounds_\xi$. It is related to the R-charge $y$ via $q = 2y +
  c$ for some constant $c$ that is different for each multiplet. See
  the various tables in section \ref{sec:kaluza-klein-analysis}.}
$q$ along the Reeb vector $\xi$ via $\Delta_0 \ge q (q + 6)$,
with equality if and only if the eigenfunction lifts to a holomorphic
function on the Calabi-Yau cone.  Equivalently, the eigenfunction is holomorphic with
respect to the tangential Cauchy-Riemann operator $\tgdb$. The
Lichnerowicz obstruction imposes $q \geq 1$, with equality if and only if the
manifold is isometric to $S^7$ \cite{Gauntlett:2006vf}.

If the scalar eigenfunction is holomorphic, a number of wave-functions vanish and
the graviton multiplet shortens. Thus, each element 
$f \in H^0(X,\mathcal{O}_X)$ defines a short graviton multiplet. In each
multiplet, the scalar eigenfunction $f$ is the wave-function of the graviton with energy $y + 3$, spin 2, and R-charge $y.$
However, the primary has energy $y + 2$, spin 1, and R-charge $y.$
Within the superconformal multiplet only the mode $\chi^{+}$ contributes to the index.  It has energy
$y + 5/2$ and spin $3/2,$ so the net contribution to the index is
$(-1) t^{y + 4}$. Since $H^0(X,\mathcal{O}_X) \cong H^0(X,\wedge^3
\Omega'_X)$, we can express the contribution of the short graviton
multiplets in terms of either cohomology group. Note however that
the map between the two cohomology groups involves the holomorphic volume form, which carries R-charge $2$.

\subsection{The Short Gravitino Multiplet}

The two gravitino multiplets $\chi^+$ and $\chi^-$ can be constructed
from one-form eigenmodes on the Sasaki-Einstein manifold. These one-forms are
the wave-functions of the vector fields $A$ and $W$ respectively. The
multiplets are listed in tables \ref{tab:gravitino_mul_1} and
\ref{tab:gravitino_mul_2}. By comparing the action of $\tgd\tgdb$ and
$J\wedge$ on one-forms, we conjecture that there is a holomorphy bound
$\Delta_1 \geq q(q+4)$, where $y + 1 = 2q$. This is equivalent to
the standard unitarity condition $E_0 \geq y + \frac{3}{2}$. Its saturation implies
that $\chi^+$ shortens. Thus every element of $H^0(X,
\Omega^\prime_X)$ defines a short gravitino multiplet.
The contribution to the index, $t^{y + 3}$, comes from the mode
$A$ with R-charge $y + 1$, spin $1$, and energy $y + 2$.

\subsection{The Short Vector and Betti Multiplets}
\label{sec:short-vector-and-betti-multiplets}
The short vector multiplet $A$ arises from holomorphic $(2,0)$
forms. Since the wave-function of the primary mode is a scalar, it is convenient to construct all wave-functions in terms of
a scalar, as we did in the case of the graviton. The primary has energy
and R-charge $E_0$ and $y$.
The holomorphic volume form $\Omega$ maps scalars $f$ to $(2,0)$ forms
by $\tgdb f \lrcorner \Omega$.  Since $\Omega$ carries R-charge 2, the
R-charge of the two-form is $y + 2$. The holomorphy bound on
two-forms can then be expressed in terms of the $\Delta_0$ eigenvalue
of $f$. Accommodating for the shifted charge, one finds $\Delta_0 = 4
E_0 (E_0 + 3) \geq
4 (y + 1)(y+4)$. The inequality is saturated if the two-form is holomorphic. Then
$E_0 = y + 1$ and the multiplet shortens since a number of
wave-functions disappear. 
Hence, the elements of
$H^0(X, \wedge^2 \Omega^\prime_X)$ correspond to
short vector multiplets $A$.
For further details see table \ref{tab:vector_mul_long_A} and
the discussion in section \ref{sec:short-multiplets}.

Finally we turn to the vector multiplet $Z$, shown in table
\ref{tab:vector_mul_Z}. It is constructed from
primitive\footnote{Recall that a form $\omega$ is primitive if it is
  annihilated by the adjoint of the Lefschetz operator -- $J \lrcorner
  \omega = 0$.}
$(1,1)$ forms and shortens when these forms are holomorphic and the
bound $\Delta_1 \geq q(q+2)$ is saturated. The primary is a scalar
with energy $E_0,$ R-charge $y,$ and a three-form wave-function.  As argued
in appendix \ref{sec:lefschetz-decomposition}, the primitive forms fill
out the cohomology group $H^{1,1}_\tgdb$. Therefore the elements of
$H^1(X, \Omega^\prime_X)$ correspond to the short vector multiplets $Z$.
In section \ref{sec:multiplets} we will show that $H_{dR}^2(Y) \subseteq H^1(X, \Omega^\prime_X).$
Therefore we find $b_2(Y)$ Betti multiplets \cite{DAuria:1984vv, DAuria:1984vy,Fabbri:1999hw, Merlatti:2000ed,Benishti:2010jn}
and possibly additional charged modes
\footnote{It would be interesting to find or exclude the possibility
  of additional charged modes.}.

\subsection{Hypermultiplets}
There are two different sources for hypermultiplets coming from Kaluza-Klein reduction.
The first arises from holomorphic scalar eigenfunctions.  When the
scalar eigenfunctions are holomorphic, the long vector multiplets $A$
shorten into hypermultiplets. Thus, each element of $H^0(X,
\mathcal{O}_X)$ defines a hypermultiplet. The scalar eigenfunction is the
wave-function of the scalar primary, which has energy and charge
$y$. The hypermultiplets contribute $t^{y}$ to the
index.

A second source for hypermultiplets are primitive, holomorphic
$(2,1)$ and $(1,2)$ forms on the Sasaki-Einstein manifold (table
\ref{tab:hyper_multiplet_2-1-form}). Here, the holomorphy bound is
$\Delta \geq q^2$. Their contribution to the index is also $t^{y}$. It
follows from our arguments in appendix
\ref{sec:lefschetz-decomposition} that all elements of $H_\tgdb^{2,1}$
and $H^{1,2}_\tgdb$ are primitive. 

The multiplets corresponding to the groups $H^{2,1}_\tgdb$ and $H^{1,2}_\tgdb$ contribute differently to the index.  Multiplets corresponding to
$H^{1,2}_\tgdb$ have a $(1,2)$ form as their primary and an accompanying Lichnerowicz mode.  However the situation is reversed for multiplets corresponding to
$H^{2,1}_\tgdb$; their primary is a Lichnerowicz mode and they have an accompanying $(2,1)$ form.
Thus, the two modes contribute differently to the index.
Hypermultiplets appear in complex conjugate pairs.  
For each hypermultiplet containing a $(1,2)$ form, there is another hypermultiplet with a $(2,1)$ form of opposite charge.
If a hypermultipet contributes to the index, its conjugate does not necessarily have to contribute as well.
One has to be careful to avoid overcounting of modes with charge zero.
We propose that the zero-charge sector should not contribute to either
hypermultiplet, since the relevant three-forms carry zero
$U(1)$-charge and are thus closed. It is likely that they are gauge modes.

\section{Cohomology Calculations}\label{cohomology}
In this section, we explain how to evaluate the trace over the cohomology groups contributing to the index.  We consider both toric and Fano manifolds.  We find several remarkable cancellations that simplify the index.
\subsection{Toric Calabi-Yau Varieties}\label{sec:toricsimplification}
Let us consider a toric Calabi-Yau cone $X$. 
In this case, there are one superconformal R-symmetry and three mesonic flavor symmetries. We will take a new basis of these symmetries such that the exponentiated chemical potentials are given by $x_1,x_2,x_3, x_4$ with $t^2=x_1x_2x_3x_4$.
Then each holomorphic function $f$ has integer charges $\bq=(q_1,q_2,q_3,q_4)$ under the four isometries, and contributes  
$\bx^\bq=x_1{}^{q_1}x_2{}^{q_2} x_3{}^{q_3}x_4{}^{q_4}$  to the index. The charges form a cone $M \subset \bZ^4$ and \begin{equation}
\Tr \bx^\bq \bigm|H^0(X,\cO_X)= \sum_{\bq\in M} \bx^\bq .
\end{equation} 
Since $\wedge^3 \Omega'_X$ carries R-charge two,
\begin{equation}
\Tr \bx^\bq \bigm|H^0(X,\wedge^3 \Omega'_X)= \sum_{\bq\in M} \bx^{\bq+(1,1,1,1)}.
\end{equation}
The groups $H^{\ge 1}(X,\wedge^k \Omega'_X) $ vanish. The characters of the ordinary and reduced differential forms are
\begin{align}
\Tr \bx^\bq \bigm| H^0(X,\wedge^k\Omega_X)
& =\sum_{\bq\in M} \tilde{n}^{k}_{\bq} \bx^\bq \\
\Tr \bx^\bq \bigm| H^0(X,\wedge^k\Omega'_X)
& =\sum_{\bq\in M} n^{k}_{\bq} \bx^\bq. \\
\end{align}
These characters are  \cite{Cox, MR495499}
\begin{equation}
\tilde{n}^k_q=\begin{cases}
0 & \text{if $q$ is on a vertex  of $M$},\\
\binom{m}{k} & \text{if $q$ is on a $m$-dimensional facet of $M$},\\
\end{cases}
\end{equation}
\begin{equation}
n_q^k =\begin{cases}
0 & \text{if $q$ is on a vertex  of $M$},\\
\binom{m-1}{k} & \text{if $q$ is on a $m$-dimensional facet of $M$}.\\
\end{cases}
\end{equation}
Let $M^{(m)}$ be the set of points of $M$ that are contained in a $m$-dimensional facet and let $M^{\circ (m)}$ be the set of points of $M$ that are contained in the interior of a $m$-dimensional facet.
Since all of the points of $M$ are in the interior of some facet, we have
\begin{equation}
\sum_{k = 1}^{\text{dim} X} \sum_{\bq \in M^{\circ (m)}} \bx^\bq =\sum_{\bq \in M^{(\text{dim} X)}} \bx^\bq= \sum_{\bq \in M^{\circ(m)}} \bx^{\bq+(1,1,1,1)}.
\end{equation}
Using these identities, we can write the index as
\begin{equation}
\label{eq:toricsci}
1 + \cI_{s.t.} = - b_2(Y) \bx^{(1,1,1,1)} + \sum_{\bq \in M^{(2)}} \left(  \bx^\bq - \bx^{\bq+(1,1,1,1)} \right) -  \sum_{\bq \in M^{(1)}}  \bx^{\bq+(1,1,1,1)}.
\end{equation}
Since we computed the cohomology of the singular toric variety, we missed the contribution of the of $b_2(Y)$ Betti multiplets which only appear on the resolved geometry, so we must add their contribution to the index.
\subsection{Toric Examples}\label{sec:toricex}
\begin{figure}[htbp]
\begin{center}
\includegraphics[width=6cm]{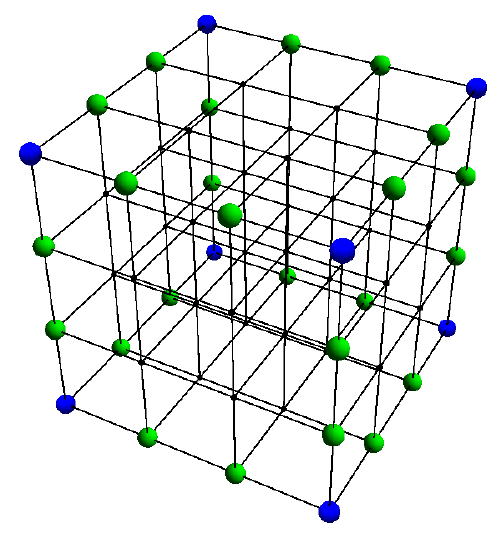}
\caption{Contributions to $M^{\circ (1)}$ and $M^{\circ (2)}$ for $Q^{1,1,1}$ with $l=3$ are colored green and blue respectively.}
\label{fig:Q111}
\end{center}
\end{figure}

We now illustrate our general formula for toric cohomologies in two simple examples.  Our first example is the cone over $Q^{1,1,1}.$  The R-symmetry group is $SU(2)^3 \times U(1)_R.$  The holomorphic functions of R-charge $t^{l}$ for integer $l$ are in one-to-one correspondence with the lattice points of the cube $[0,l]^3.$
The flavor fugacity transforms in the $[l,l,l]$ representation of $SU(2)^3$ of dimension $(l+1)^3$.
The lattice points are shown in figure \ref{fig:Q111} for $l=3$.  In table \ref{tab:Q111toric} we list the contributions to $M^{\circ (m)}$ for small $l.$  Using this data, we the find the contributions to the index listed in table \ref{tab:Q111sci}.  Summing all the contributions to the index, we find
\begin{equation}
\label{eq:Q111}
\cI_{s.t.}(Q^{1,1,1};t) = 8t + \frac{16 t^2}{1 - t}.
\end{equation}
\begin{table}[htdp]
\begin{center}
\begin{tabular}{|l|c|cccccc|}
\hline
Contribution & Fugacity							& 1 & $t$ & $t^2$ & $t^3$ & $t^4$ & $\dots$ \\
\hline 
$\sum_{\bq \in M^{(2)}}$ & $ \bx^\bq $			& 1  & 8  & 20 & 32 & 44 & $\dots$ \\ 
$ \sum_{\bq \in M^{(2)}}$ & $ - \bx^{\bq+(1,1,1,1)}$	& 0  & 0  &   1  &  8  & 20 & $\dots$\\
$ \sum_{\bq \in M^{(1)}}$ & $ - \bx^{\bq+(1,1,1,1)}$	&  0  & 0  &   1 &  8  & 8 & $\dots$\\
$ b_2(Y)$		& $-\bx^{(1,1,1,1)}$			& 0   & 0  &   2 &  0  & 0& $\dots$ \\
\hline
$1+\cI_{s.t.}$		& $(-1)^F \bx^{\epsilon + j_3}$	& 1  	& 8  &  16 & 16 & 16 & $\dots$\\
\hline
\end{tabular}
\end{center}
\caption{Contributions to the index of $Q^{1,1,1}$.}
\label{tab:Q111sci}
\end{table}%

\begin{table}[htdp]
\begin{center}
\begin{tabular}{c|cccc}
$l$ & $M^{\circ (1)}$ & $M^{\circ (2)}$ & $M^{\circ (3)}$ & $M^{\circ (4)}$ \\
\hline
0 & 1 & 0 & 0 & 0 \\
1 & 8 & 0 & 0 & 0 \\
2 & 8 & 12 & 6 & 1 \\
3 & 8 & 24 & 24 & 8 \\
4 & 8 & 36 & 54 & 27 \\
5 & 8 & 48 & 96 & 64 \\
\end{tabular}
\end{center}
\caption{Toric data for $Q^{1,1,1}$.}
\label{tab:Q111toric}
\end{table}
Our second example is the seven-sphere $S^7.$  The cone over $S^7$ is the four-dimensional complex space $\mathbb{C}^4$.  The ring of holomorphic functions is $\mathbb{C}[x_1,x_2,x_3,x_4].$
Monomials with fixed degree $l$ have R-charge $l/2$ and correspond to the lattice points of a tetrahedron with $(l+1)$ lattice points on each side.  We list the contributions to the index in table \ref{tab:S7sci}.  Summing the contributions to the index, we find 
\begin{equation}
\label{eq:S7}
\cI_{s.t.}(S^{7};t) =  4t^{1/2} + 10t + 16 t^{3/2} \frac{20 t^2}{1 - t^{1/2}}.
\end{equation}
\begin{table}[htdp]
\begin{center}
\begin{tabular}{|l|c|cccccccc|}
\hline
Contribution & Fugacity							& 1 & $t^{1/2}$ & $t$ & $t^{3/2}$ & $t^2$ & $t^{5/2}$ & $t^3$ & $\dots$ \\
\hline 
$\sum_{\bq \in M^{(2)}}$ & $ \bx^\bq $			& 1  & 4  & 10 & 16 & 22 & 28 & 34 & $\dots$ \\ 
$ \sum_{\bq \in M^{(2)}}$ & $ - \bx^{\bq+(1,1,1,1)}$	& 0  & 0  &   0  &  0  & 1 & 4 & 10 & $\dots$\\
$ \sum_{\bq \in M^{(1)}}$ & $ - \bx^{\bq+(1,1,1,1)}$	&  0  & 0  &   0 &  0  & 1 & 4 & 4 & $\dots$\\
$ b_2(Y)$		& $-\bx^{(1,1,1,1)}$			& 0   & 0  &   0 &  0  & 0& 0 & 0 & $\dots$ \\
\hline
$1+\cI_{s.t.}$		& $(-1)^F \bx^{\epsilon + j_3}$	& 1  	& 4  &  10 & 16 & 20 & 20 & 20 & $\dots$\\
\hline
\end{tabular}
\end{center}
\caption{Contributions to the index of $S^{7}$.}
\label{tab:S7sci}
\end{table}%

\subsection{Cones over Fano Varieties}\label{sec:fano}
Our results take an especially nice form for cones over Fano threefolds.  A smooth projective variety is called Fano if its anti-canonical divisor $-K_V \cong \det T_V$ is ample.  We now introduce several invariants of Fano threefolds $V$ and express them in terms of the Chern classes $c_{i}(V).$  The degree of $V$ is $d = -K_V^3 = \int_{V} c_1^3$ and the Euler characteristic $\chi(V) =  \int_{V} c_3  = 2(1+ b_2 - b_3).$  We will simplify many expressions using $\int_{V} c_1^2 c_2 = 24$
\footnote{This relation can be derived using the Hirzebruch-Riemann-Roch theorem to calculate $\chi(\mathcal{O}_V) = 1$.}.

The index $\text{ind}(V)$ is the largest integer $r$ such that there exists a divisor $H$ such that $rH \cong -K_V.$  We call $H$ the fundamental divisor.  Let $\mathcal{L}$ be the line bundle corresponding to the fundamental divisor.  If $V$ has a K\"ahler-Einstein metric then the total space of the fibration $\mathcal{L} \rightarrow V$ is a Calabi-Yau cone.  The unit circle bundle in $\mathcal{L}$ is a Sasaki-Einstein manifold \cite{Martelli:2006yb, Gauntlett:2006vf}.  For simplicity we will restrict our attention to this case.

Fano threefolds are classified by the work of Fano, Iskovskikh, Shokurov, Mori, and Mukai.  The index of an n-dimensional Fano variety satisfies $\text{ind}(V) \le n+1.$  The complex projective space $\CP^3$ is the unique Fano threefold with index 4.  Similarly the quadric hypersurface $Q$ in $\CP^4$ is the unique Fano threefold with index 3.  There are 105 deformation families of Fano threefolds and  they are described in \cite{MR1668575}.
We list a few of the simplest Fano threefolds in table \ref{tab:SElist}.


\begin{table}[htdp]
\begin{center}
\begin{tabular}{|c|c|c|c|c|c|}
\hline
$SE_7$		& Fano							& $b_2(Y)$	& $\chi(V)$ 	& $ind(V)$ & $-K_V^3$ \\
\hline
$S^7$		& $\CP^3$						& 0 			& 4			& 4 		& 64 \\
$M^{1,1,1}$	& $\CP^2 \times \CP^1$				& 1			& 6			& 1		& 54 \\
$Q^{1,1,1}$	& $\CP^1 \times \CP^1 \times \CP^1$	& 2			& 8			& 2 		& 48 \\
$V_{5,2}$		& $Q$							& 0			& 4			& 3 		& 54 \\
\hline
\end{tabular}
\end{center}
\caption{Some Fano threefolds and their Sasaki-Einstein cones.}
\label{tab:SElist}
\end{table}%

\subsection{Twisted Cohomology of Fano Varities}
In this section we compute the cohomology groups contributing to the superconformal index for cones over Fano threefolds.  Since $V$ is Fano, its anti-canonical divisor $-K_V$ is ample and the cohomology groups $H^{i \ge 1}(V, \mathcal{L}^j)$ vanish by the Kodaira vanishing theorem.
The first contribution to the index is the Hilbert series.  The Hilbert series of the cone can be computed using the Hirzebruch-Riemann-Roch theorem since all higher cohomology groups vanish.
We compute the Hilbert series using
\begin{align}
C(t, V)  & =  \sum_{j \ge 0} t^{2j/r} \chi(V, \mathcal{L}^j) \\
&  = \sum_{j \ge 0} t^{2j/r} \int e^{-j c_1(\mathcal{L})} \cdot Todd(V),
\end{align}
where $r$ is the Fano-index.
The Todd class  can be expressed in terms of the Chern classes of the tangent bundle $TV$ as
\begin{equation}
Td(V) = 1 + \frac{c_1(TV)}{2} + \frac{c_1(TV)^2 + c_2(TV)}{12} + \frac{c_1(TV) c_2(TV)}{24}.
\end{equation}
For Fano threefolds of index 1, $\mathcal{L} = -K_V$
and
\begin{equation}
e^{- c_1(\mathcal{L})} = 1 + c_1(K_V)+ c_1(K_V)^2/2 + \dots
\end{equation}
Combing these ingredients, we find that the Hilbert series is
 \begin{align}
 C(t,V) &  = \frac{1 + (d/2-1) t^2 + (d/2-1) t^4+t^6}{(1-t^2)^4}
 \end{align}
where $d$ is the degree.
We can similarly compute the other contributions to the superconformal index using the Hirzebruch-Riemann-Roch theorem.  
Starting with our general formula for the index in equation \eqref{aab}, we can organize the contributions to the index in terms of the characters
\begin{equation}
C^j(t, V) =  \sum_n t^n \chi(V,\wedge^j \Omega_V \otimes \mathcal{L}^n).
\end{equation}
Using Kodaira-Nakano vanishing theorem, $H^{p} (V,\wedge^q \Omega_V \otimes \mathcal{L}^n) = 0$ for $p + q > \text{dim}(V),$ we find that
the superconformal index is
\begin{equation}
1 + \cI_{s.t.}(t) = C^0 - C^2 + t^2 (C^1 - C^3).
\end{equation}
For Fano threefolds of index $\ind(V) = 1$, $\mathcal{L} = -K_V$, and we find that the superconformal index is
\begin{equation}
\cI_{s.t.}(t) =  \frac{(24 - \chi(V))t^2}{1 - t^2}.
\end{equation}
For Fano threefolds of index $\ind(V) = 2$:
\begin{equation}
\cI_{s.t.}(t) =  (12 -  \chi(V)/2) t +  \frac{(24 - \chi(V))t^2}{1 - t}.
\end{equation}
We can also compute the index for $\CP^3$ and the quadric $Q$ with all flavor fugacities set to one.  We find that the superconformal indices are
\begin{align}
\cI_{s.t.}(\CP^3;t) & =  4t^{1/2} + 10t + 16 t^{3/2} + \frac{20 t^2}{1 - t^{1/2}}, \text{ and}  \label{eq:S7f} \\
\cI_{s.t.}(Q; t) & =  6t^{2/3} + 14 t^{4/3} + \frac{20 t^2}{1 - t^{2/3}}. \label{eq:v52} \\
\end{align}
In the next section we will derive the superconformal indices for the cones over $\CP^3$ and the quadric, $Q$, refined by their flavor fugacities.
\subsection{The Seven-Sphere $S^7$}
The seven-sphere $S^7$ is the unit circle bundle over $\CP^3.$  Its isometry group is $SU(4) \times U(1)_R.$
Using the twisted cohomology groups of complex projective space listed in appendix \ref{cohomology-appendix}, we find the short multiplets in supergravity
and list them in table \ref{tab:S7}.
\begin{table}[htdp]
\begin{center}
\begin{tabular}{|l|c|l|l|l|c|}
\hline
Multiplet 			& $j_3$	& Primary $y$ 			& Multiplicity & $SU(4)_{U(1)_R}$ & Index  \\
\hline
short graviton 		& 1	&	$\ell/2-2$				& $H^0(X,\wedge^3 \Omega'_X)$		& $[\ell-4,0,0]_{\ell/2}$	& $-t^{\ell/2 + 2}$ \\
short gravitino 		& 1/2&	$\ell/2-1$				& $H^0(X, \Omega'_X)$				& $[\ell-2,1,0]_{\ell/2}$	& $t^{\ell/2 + 2}$ \\
short vector $A$ 		& 0	&	$\ell/2-2$				& $H^0(X,\wedge^2 \Omega'_X)$		& $[\ell-3,0,1]_{\ell/2}$	& $-t^{\ell/2}$\\
hyper 			& 0	&	$\ell/2$				& $H^0(X,\mathcal{O}_X)$			& $[\ell,0,0]_{\ell/2}$		& $t^{\ell/2}$\\
\hline
\end{tabular}
\end{center}
\caption{Short multiplets for $AdS_4 \times S^7$ and their contribution to the index.  Some short multiplets can become massless for small $\ell.$}
\label{tab:S7}
\end{table}
Summing all contributions to the index, we find
\begin{equation}
1 + \cI_{s.t.}(S^7; t) = \sum_{n \ge 0 } (1 - t^4) t^{n/2} \chi^{SU(4)}_{[n,0,0]} - t^{n/2 + 3/2} \chi^{SU(4)}_{[n,0,1]} + t^{n/2+3} \chi^{SU(4)}_{[n,1,0]}.
\end{equation}
This is in complete agreement with the original derivation of the supergravity index \cite{Bhattacharya:2008zy} and the field theory index \cite{Bhattacharya:2008bja, Kim:2009wb}.
\subsection{The Stiefel Manifold $V_{5,2}$}
The Stiefel manifold $V_{5,2} = SO(5)/SO(3)$ is the unit circle bundle over the quadric $Q \in \CP^4.$
Its isometry group is $SO(5) \times U(1)_R$.
Using the twisted cohomology groups of the quadric listed in appendix \ref{cohomology-appendix}, we find the short multiplets in supergravity
and list them in table \ref{tab:V52}.  We write $SO(5)$ representations in terms of the Dynkin labels $[\lambda_1, \lambda_2]$ of their highest weight state.
\footnote{If $\lambda_2$ is odd, the representation $[\lambda_1, \lambda_2]$ is spinor.  If $\lambda_2$ is even, the representation is tensor and the corresponding Young tableaux has $\lambda_1$ height one columns and $\lambda_2/2$ height two columns.}
The Kaluza-Klein spectrum for $V_{5,2}$ was first derived in \cite{Ceresole:1999zg}
\footnote{Additional long multiplets were found in \cite{Pilch:2013fk}.}.
We find that one of the short gravitino multiplets listed in \cite{Ceresole:1999zg} is extraneous.  The non-trivial cohomology group $H^1(Q, \Omega^2(1)) \cong \bC$ implies the existence of a short hypermultiplet also missing from \cite{Ceresole:1999zg}.  This hypermultiplet and its anti-holomorphic partner were previously found in \cite{Cassani:2012pj}.  
\begin{table}[htdp]
\begin{center}
\begin{tabular}{|l|c|l|l|l|c|}
\hline
Multiplet 			& $j_3$	& Primary $y$ 						& Multiplicity & $SO(5)_{U(1)_R}$ & Index  \\
\hline
short graviton 		& 1		&$2\ell/3-2$				& $H^0(X,\wedge^3 \Omega'_X)$		& $[\ell-3]_{2\ell/3}$		& $-t^{2\ell/3 + 2}$ \\
short gravitino 		& 1/2		&$2\ell/3-1$				& $H^0(X, \Omega'_X)$				& $[\ell-2,2]_{2\ell/3}$	& $t^{2\ell/3 + 2}$ \\
short vector $A$	& 0		&$2\ell/3-2$				& $H^0(X,\wedge^2 \Omega'_X)$		& $[\ell-3,2]_{2\ell/3}$	& $-t^{2\ell/3}$\\
hyper 			& 0		&$2\ell/3$				& $H^0(X,\mathcal{O}_X)$			& $[\ell,0]_{2\ell/3}$		& $t^{2\ell/3}$\\
hyper	 		& 0		& $2\ell/3$			& $H^1(X,\wedge^2 \Omega'_X)$		& $[0,0]_{2/3}$ 			& $t^{2/3}$\\
\hline
\end{tabular}
\end{center}
\caption{Short multiplets for $V_{5,2}$.  Here the $U(1)_R$ charge is $y.$}
\label{tab:V52}
\end{table}
Summing all contributions, the gravity superconformal index is
\begin{equation}
\label{eq:v52ri}
1 + \cI_{s.t.}(V_{5,2}; t) = t^{2/3} + \sum_{n \ge 0 }(1-t^4) t^{2/3 n} \chi^{SO(5)}_{[n,0]} - (1-t^{4/3}) \chi^{SO(5)}_{[n,2]} t^{2 + 2/3 n},
\end{equation}
which refines the index in equation \eqref{eq:v52} by the $SO(5)$ flavor fugacities.

\begin{figure}[h]
\begin{center}
\begin{tikzpicture}[scale=0.8] 
\tikzset{ 
my loop/.style={->,to path={ 
.. controls +(-45:2) and +(45:2) .. (\tikztotarget) \tikztonodes}}
} 
\tikzset{ 
my loopb/.style={->,to path={ 
.. controls +(135:2) and +(225:2) .. (\tikztotarget) \tikztonodes}}
} 
\path (0,0) node[draw,shape=circle] (v0) {$1$}; 
\path (0:4cm) node[draw,shape=circle] (v1) {$2$}; 
\path[->,thick, bend right = 20] (v1) edge node[above] {$$} (v0);
\path[->,thick, bend right = 40] (v1) edge node[above] {$A_1, A_2$} (v0);
\path[->, thick, bend right = 40] (v0) edge node[below] {$B_1, B_2$} (v1);
\path[->, thick, bend right = 20] (v0) edge node[below] {$$} (v1);
\path[thick] (v1) edge[my loop] node[right] {$\Phi_2$} (v1);
\path[thick] (v0) edge[my loopb] node[left] {$\Phi_1$} (v0);
\end{tikzpicture}
\caption{Quiver for $V_{5,2}$} 
\label{fig:quiver}
\end{center}
\end{figure}
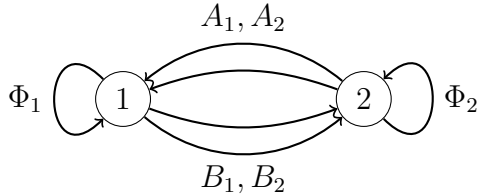

The field theory dual to M2-branes at the cone over $V_{5,2}$ was recently proposed to be a $U(N) \times U(N)$ gauge theory with Chern-Simons interactions \cite{Martelli:2009ga}.  The bi-fundamental and adjoint matter content is shown as a quiver in figure \ref{fig:quiver}.  The manifest global symmetry is $SU(2) \times U(1)_B \times U(1)_R.$  The pairs of bifundamental fields $(A_1, A_2)$ and $(B_1, B_2)$ transform as doublets under the global $SU(2)$ symmetry. 
The field theory index was computed in \cite{Imamura:2011uj} up to terms of order $t^2$.  The gravity index we compute in equation \eqref {eq:v52ri} exactly matches the known field theory terms.
However, a few more terms are desirable, since the leading contribution of the short gravitino multiplet in \cite{Ceresole:1999zg} is at order $t^{10/3}.$  While it is possible to directly extend the field theory computation to arbitrary fixed order, we instead compute the index to all orders in $t$ in the zero-monopole sector for simplicity.
Since the each pair of bi-fundamental fields transforms as a doublet under the $SU(2)$ global symmetry, we assign the flavor fugacity  $\chi^{SU(2)}_{1/2}(z) = z^{1/2} + z^{-1/2}$ to each pair.
Let $M_{Q}(t,z)$ be the adjacency matrix of the $V_{5,2}$ quiver weighted by the R-charges and flavor fugacities,
\begin{equation}
M_{Q}(t,z) =
\begin{pmatrix}
t^{2/3} & t^{2/3} \chi^{SU(2)}_{1/2}(z) \\
t^{2/3} \chi^{SU(2)}_{1/2}(z) & t^{2/3}
\end{pmatrix}
,
\end{equation}
and let $\chi_Q(t,z)$ be the matrix constructed from the single-letter indices.  In terms of the adjacency matrix, $\chi_Q(t,z)$ has the following simple form,
\begin{equation}
\chi_Q(t,z) = \frac{1}{(1-t^2)} \left(1 - M_Q(t,z) + t^2 M_Q^{T}(t^{-1},z^{-1}) - t^2 \right).
\end{equation}
The multi-trace contribution to the index from the letters is
\begin{equation}
I^{(0)} = \prod_{n \ge 1} \frac{1}{\det(\chi_Q(t^n,z^n))}.
\end{equation}
Using the plethystic logarithm we can extract the single-trace index in the zero-monopole sector
\begin{equation}
\mathcal{I}_{s.t.}^{(0)} = (1-t^2) PE^{-1}[I^{(0)}].
\end{equation}
We find that the single-trace index in the zero-monopole sector is
\begin{equation}
\mathcal{I}_{s.t.}^{(0)} = 2 t^{2/3} + (2 + z + z^{-1}) t^{4/3} + \sum_{n \ge 2}  \left( z^n + z^{-n} \right)(1 - t^{2/3})  t^{4/3 n}.
\end{equation}
In supergravity, the zero-monopole sector corresponds to the states that remain in the orbifold $V_{5,2} /  \mathbb{Z}_k$ in the large $k$ limit.
Equivalently, these are the states that are singlets under $SU(2)_L$ in the decomposition $SO(5) \rightarrow SU(2)_L \times SU(2)_R.$  Using the branching rules for $SO(5),$ we verify that the supergravity index \eqref{eq:v52ri} exactly matches the field theory index in the zero-monopole sector. 
\section{The Supergravity Analysis}
\label{sec:kaluza-klein-analysis}

In this section we turn to the analysis of eleven-dimensional
supergravity on $AdS_4 \times SE_7$ manifolds. Standard results in
gauge/string duality relate the operator spectrum of the dual gauge
theory to the spectrum of various differential operators on the
Sasaki-Einstein manifold. For reference, we give this dictionary in
tables \ref{tab:mass-relations} and
\ref{tab:scaling_dim_from_ads_mass}.

In section \ref{sec:kaluza-klein-spectrum} we determine the spectrum
of $\Delta$ and $Q = \star d$. We leave the analysis of
fermionic and Lichnerowicz eigenmodes to a future work. Section
\ref{sec:multiplets} explains how to arrange the various modes into
multiplets.

\subsection{The Kaluza-Klein Spectrum}
\label{sec:kaluza-klein-spectrum}

\begin{table}[hbt]
  \centering
  \begin{equation*}
    \begin{array}{|ccccc|}
      \hline
      \text{Name} & \text{Degree} & \Delta, Q & \text{Charge} & \\
      \hline
      f^{\lbrack 0; q\rbrack} & 0 & \delta & q & c, \star, \bar{\star}\\
      \hline
      df^{\lbrack 0; q\rbrack} & 1 & \delta & q  & \\
      f^{\lbrack 1; q; -\rbrack} & 1 & \delta + 6 - 2 \sqrt{\delta + 9} & q & \bar{\star}
      \\
      f^{\lbrack 1; q; +\rbrack} & 1 & \delta + 6 + 2 \sqrt{\delta + 9} & q & c, \star \\
      \hline
      f^{\lbrack 2; q; a\rbrack}  & 2 & \delta + 8 & q & \star, \bar{\star} \\
      f^{\lbrack 2; q; b\rbrack}  & 2 & \delta + 8 & q & \\
      d f^{\lbrack 1; q; -\rbrack} & 2 & \delta + 6 - 2 \sqrt{\delta + 9} & q & \\
      d f^{\lbrack 1; q; +\rbrack} & 2 & \delta + 6 + 2 \sqrt{\delta + 9} & q & \\
      f^{\lbrack 2; q+4 \rbrack} & 2 & \delta + 8 & q + 4 & \bar{\star} \\
      f^{\lbrack 2; q-4\rbrack} & 2 & \delta + 8 & q-4 & \star \\
      \hline
      f^{\lbrack 3; q; -\rbrack} & 3 & 1 - \sqrt{\delta + 9} & q & \bar{\star} \\
      f^{\lbrack 3; q; +\rbrack} & 3 & 1 + \sqrt{\delta + 9} & q  & c, \star \\
      f^{\lbrack 3; q+4; -\rbrack} & 3 & 1 - \sqrt{\delta + 9} & q+4 &
      \bar{\star} \\
      f^{\lbrack 3; q+4; +\rbrack} & 3 & 1 + \sqrt{\delta + 9} & q+4 &
      c, \star, \bar{\star} \\
      f^{\lbrack 3; q-4; -\rbrack} & 3 & 1 - \sqrt{\delta + 9} & q-4 &
      \star, \bar{\star} \\
      f^{\lbrack 3; q-4; +\rbrack} & 3 & 1 + \sqrt{\delta + 9} & q-4 &
      c, \star \\
      \hline
    \end{array}
  \end{equation*}
  \caption{Wave functions $f^{\lbrack p;q;X \rbrack}$ derived from a
    scalar $f$. $p$ refers to the form degree, $q$ to its charge, $X$
    denotes any additional labels. Non-gauge modes that remain
  for holomorphic $f$ -- i.e.~$\tgdb f = 0$ -- are labeled with
  ``$\star$''. ``$\bar{\star}$'' marks the anti-holomorphic case. The
  modes with constant $f$ are labeled ``$c$''. Neither of $f^{\lbrack
    2; q; a, b\rbrack}$ vanishes when $f$ is (anti-) holomorphic, yet
  they coincide. We label this mode $f^{\lbrack 2; q; a \rbrack}$.}
  \label{tab:wave-functions_from_scalars}
\end{table}

\begin{table}[hbt]
  \centering
  \begin{equation*}
    \begin{array}{|ccccc|}
      \hline
      \text{Name} & \text{Degree} & \Delta, Q & \text{Charge} & \\
      \hline
      \sigma^{\lbrack 1; q\rbrack} & 1 & \delta & q & \star \\
      \sigma^{\lbrack 1; q-4\rbrack} & 1 & \delta & q-4 & \star \\
      \hline
      \sigma^{\lbrack 2; q; -\rbrack} & 2 & \delta + 4 - 2
      \sqrt{\delta + 4} & q & \\
      \sigma^{\lbrack 2; q; +\rbrack} & 2 & \delta + 4 + 2
      \sqrt{\delta + 4} & q & \star \\
      \sigma^{\lbrack 2; q-4; -\rbrack} & 2 & \delta + 4 - 2
      \sqrt{\delta + 4} & q -4 & \star \\
      \sigma^{\lbrack 2; q-4; +\rbrack} & 2 & \delta + 4 + 2
      \sqrt{\delta + 4} & q - 4 & \star \\
      \hline
      \sigma^{\lbrack 3; q; -\rbrack} & 3 & - \sqrt{\delta + 4} & q &
      \\
      \sigma^{\lbrack 3; q; +\rbrack} & 3 & \sqrt{\delta + 4} & q &
      \star \\
      \sigma^{\lbrack 3; q-4; -\rbrack} & 3 & - \sqrt{\delta + 4}
            & q-4 & \star \\
      \sigma^{\lbrack 3; q-4; +\rbrack} & 3 & \sqrt{\delta + 4} & q-4
      & \star \\
      \hline
    \end{array}
  \end{equation*}
  \caption{Wave functions $\sigma^{\lbrack p;q;X \rbrack}$ derived from a
    $(1,0)$-form $\sigma$. Again, holomorphic modes are marked ``$\star$''.}
  \label{tab:wave-functions_from_one-forms}
\end{table}

\begin{table}[hbt]
  \centering
  \begin{equation*}
    \begin{array}{|ccccc|}
      \hline
      \text{Name} & \text{Degree} & \Delta, Q & \text{Charge} & \\
      \hline
      \chi^{\lbrack 2; q\rbrack} & 2 & \delta & q & \star, \square \\
      \hline
      \chi^{\lbrack 3; q; -\rbrack} & 3 & -1 - \sqrt{\delta + 1} & q &
      \star, \square \\
      \chi^{\lbrack 3; q; + \rbrack} & 3 & -1 + \sqrt{\delta + 1} & q
      & \\
      \hline
    \end{array}
  \end{equation*}
  \caption{Wave functions $\chi^{\lbrack p;q;X \rbrack}$ derived from a
    primitive $(1,1)$-form $\chi$. ``$\star$'' marks the modes remaining
    when $\chi$ is holomorphic and $\delta = q(q+2)$, ``$\square$'' those
    with $d\chi = 0$ and $\delta = q = 0$.}
  \label{tab:wave-functions_from_two-forms}
\end{table}

\begin{table}[hbt]
  \centering
  \begin{equation*}
    \begin{array}{|ccccc|}
      \hline
      \text{Name} & \text{Degree} & Q & \text{Charge} & \\
      \hline
      \zeta^{\lbrack 3; q\rbrack} & 3 & -q & q & \star \\
      \vartheta^{\lbrack 3; q \rbrack} & 3 & q & q & \star \\
      \hline
    \end{array}
  \end{equation*}
  \caption{Wave functions based on three-forms.}
  \label{tab:wave-functions_from_three-forms}
\end{table}

The Kaluza-Klein spectra of various coset manifolds have been
successfully studied using harmonic analysis. For more general
manifolds however, the problem becomes considerably more
difficult. Hence as in \cite{Eager:2012hx} our strategy is to use less
information; i.e.~we start with an eigenform of one of the
differential operators and then use the Sasaki-Einstein structure to
construct additional eigenforms. 
In other words, given an eigenfunction $f$ with $\Delta f = \delta f$,
we use interior and exterior products with the forms and operators
$\eta, J, \Omega, \tgd, \tgdb, \pounds_\xi.$
We then identify the short multiplets with the Kohn-Rossi cohomology groups
$H_{\tgdb}^{p,q}$ and the related groups $H^q(X,
\wedge^p \Omega^\prime_X)$ on the cone. Our conventions concerning
Sasaki-Einstein manifolds, Kohn-Rossi cohomology, and the various
differential operators are listed in appendix \ref{conventions}.
For details of the analysis the reader should refer to appendix
\ref{sec:details-of-supergravity-analysis}. An analysis of the
compatibility of Lefschetz decomposition with Kohn-Rossi cohomology is
performed in appendix \ref{sec:lefschetz-decomposition}.

Eigenmodes constructed from scalars $f$, $(1,0)$ forms $\sigma$,
primitive $(1,1)$ forms $\chi$, and primitive $(2,1)$ forms $\zeta$
are listed in tables \ref{tab:wave-functions_from_scalars},
\ref{tab:wave-functions_from_one-forms},
\ref{tab:wave-functions_from_two-forms}, and
\ref{tab:wave-functions_from_three-forms}. Various conditions -- most
noteably primitivity -- simply stem from the fact that we impose
orthogonality between the various modes in order to avoid
overcounting. Since the basis we use for the scalars includes $f \eta,
f J, f \eta \wedge J$, the higher forms have to be orthogonal to both
the Reeb vector and the K\"ahler form. Further conditions are
discussed in the appendix.

A crucial role is played by Kohn-Rossi holomorphic forms. 
Whenever forms or scalars are holomorphic (i.e.~annihilated by
$\tgdb$), a number of derived modes 
vanish. This will lead to multiplet shortening. The surviving modes are marked in the tables.
Moreover, given a function $f$ and
$\pounds_\xi f = \imath q f$, one can show that the Laplace operator
is bounded (see eq.~\eqref{eq:bound_f_derived})
\begin{equation}
  \Delta f \geq q (q+6) f.
\end{equation}
Equality holds if and only if $f$ is holomorphic. Our calculations in
appendix \ref{sec:details-of-supergravity-analysis} suggest that the
suitable generalization for primitive $k$-forms orthogonal to the Reeb
vector is ($k = 0, 1, 2,
3$) \begin{equation}\label{eq:p-form_holomorphy_bound}
  \Delta_k \geq q \lbrack q + (6-2k) \rbrack = 4 y_{\tgdb} 
  \lbrack y_{\tgdb} + (3-k) \rbrack.
\end{equation}
Here $y_{\tgdb} = 2q$. Again, equality holds if and only if the form is
holomorphic. This inequality prompts us to define $E_{\tgdb}$ via
\begin{equation}
  \Delta_k = 4 E_{\tgdb} \lbrack E_{\tgdb} + (3-k) \rbrack.
\end{equation}
In these variables, the bound becomes
\begin{equation}
  E_{\tgdb} \geq y_{\tgdb}.  
\end{equation}

It would be very interesting to find a further generalization of
\eqref{eq:p-form_holomorphy_bound} that holds for all $k$-forms. The
minimal modification that satisfies $\Delta J = 12 J$ and $\Delta \eta = 12 \eta$ is
\begin{equation}\label{eq:deRham-KR-Laplacian-guess}
  \Delta = 2 \Delta_\tgdb - \pounds_\xi^2 - 2 \imath (3-k)
  \pounds_\xi + 4 J \wedge J \lrcorner + 6 \eta \wedge \eta \lrcorner.
\end{equation}
$\Delta_\tgdb$ is the Kohn-Rossi Laplacian, $\Delta_\tgdb = \tgdb^\dagger \tgdb +
\tgdb \tgdb^\dagger$. In this form \eqref{eq:deRham-KR-Laplacian-guess}
begins to resemble the well-known identity for the de Rham and the
Dolbeault Laplacians on K\"ahler manifolds, $\Delta = 2\Delta_\partial
= 2\Delta_{\bar{\partial}}$.

\subsection{The Multiplets}
\label{sec:multiplets}

In this section, we will arrange the Kaluza-Klein modes into superconformal multiplets.
Substituting the Kaluza-Klein modes derived in section
\ref{sec:kaluza-klein-spectrum} into tables \ref{tab:mass-relations} and
\ref{tab:scaling_dim_from_ads_mass} yields the states of the dual
SCFT.   The superconformal
primary has energy, spin, and hypercharge $(E_0, j_3, y)$. These variables
differ from $E_{\tgdb}$ and $y_{\tgdb}$ by
constant shifts. The superconformal primaries are labeled ``$p$''.
Our
analysis here has drawn on previous results obtained in the special
case of coset manifolds. See e.g.~\cite{DAuria:1984vy}, \cite{Ceresole:1984hr},
and especially \cite{Fabbri:1999mk}. These
references also include the fermionic modes. Wave-functions for which
we have not derived an explicit expression are labeled with a
subscript ``$*$''.

\paragraph{Long multiplets}
\label{sec:long-multiplets}

\begin{table}[hbtp]
  \centering
  \begin{equation*}
    \begin{array}{|c|c|c|c|c|c|c|}
      \hline
      \text{Spin} & \text{Energy} & \text{Charge} & \text{Mass}^2
      & \text{Name} & \text{Wave-f.} &  \\
      \hline
      2 & E_0 + 1 & y & 4 (E_0-2) (E_0 + 1) & h & f^{\lbrack 0;
        q\rbrack} & c, \star \\
      \frac{3}{2} & E_0 + \frac{1}{2} & y + 1 & E_0 - 2 & \chi^+ &
      f_*^{\lbrack 3/2 \rbrack} & c, \star \\
      1 & E_0 + 2 & y & 4  E_0 (E_0+1) & W & f^{\lbrack 1; q; -\rbrack} &  \\
      1 & E_0 + 1 & y - 2 & 4 E_0 (E_0 -1) & Z & f^{\lbrack 2;
        q-4\rbrack} & \star \\
      1 & E_0 + 1 & y + 2 & 4E_0 (E_0-1) & Z & f^{\lbrack 2; q+4\rbrack} & \\
      1 & E_0 + 1 & y & 4E_0 (E_0-1) & Z & f^{\lbrack 2; q; a, b\rbrack} & \star \\
      1 & E_0 + 1 & y & 4E_0 (E_0-1) & Z & f^{\lbrack 2; q; a, b\rbrack} & \\
      1 & E_0 & y & 4 (E_0-2)(E_0 -1) & A & f^{\lbrack 1; q; +
        \rbrack} & c, \star, p \\
      0 & E_0 + 1 & y & 4E_0 (E_0-1) & \phi & f^{\lbrack 2_s; q \rbrack}_* & \\
      \hline
    \end{array}
  \end{equation*}
  \caption{The graviton multiplet. Modes surviving in the case that
    $f$ is holomorphic are labeled ``$\star$''. Then $E_0 = y +
    2$. Moreover, modes remaining if f is constant are labeled with
    ``$c$'' and satisfy $y = 0$. They fill out the massless graviton
    multiplet. The normalization is such that $E_{\tgdb} + 2 = E_0$
    and $y_{\tgdb} = y$.}
  \label{tab:graviton_multiplet}
\end{table}

\begin{table}[hbtp]
  \centering
  \begin{equation*}
    \begin{array}{|c|c|c|c|c|c|c|}
      \hline
      \text{Spin} & \text{Energy} & \text{Charge} & \text{Mass}^2
      & \text{Name} & \text{Wave-f.} &  \\
      \hline
      1 & E_0 + 1 & y & 4E_0 (E_0-1) & A & f^{\lbrack 1; q; -\rbrack}
      & m, \diamond \\
      \frac{1}{2} & E_0 + \frac{1}{2} & y + 1 & E_0 - 1 & \lambda_L
      & f^{\lbrack 1/2\rbrack}_* & \diamond \\
      0 & E_0 + 2 & y & 4E_0 (E_0+1) & \phi & f^{\lbrack 2_s; q
        \rbrack}_* & \\
      0 & E_0 + 1 & y - 2 & 4E_0 (E_0 -1) & \pi & f^{\lbrack 3; q-4; -\rbrack} & \diamond \\
      0 & E_0 + 1 & y + 2 & 4E_0 (E_0-1) & \pi & f^{\lbrack 3; q+4; -\rbrack} & \\
      0 & E_0 + 1 & y & 4E_0(E_0-1) & \pi & f^{\lbrack 3; q;
        -\rbrack} & m, \diamond \\
      0 & E_0 & y & 4(E_0-2)(E_0-1) & S & f^{\lbrack 0; q\rbrack} &
      m, \diamond, p \\
      \hline
    \end{array}
  \end{equation*}
  \caption{The vector multiplet $A$. ``$\diamond$'' marks the modes
    remaining when $\tgdb f \lrcorner \Omega$ defines a holomorphic
    $(2,0)$ form and the multiplet shortens. In this case
    $E_0=y+1$. The normalization is simple: $E_{\tgdb} = E_0$,
    $y_{\tgdb} = y$. ``$m$'' marks the massless multiplet. We
    include the fermion mode contributing to the index.}
  \label{tab:vector_mul_long_A}
\end{table}

\begin{table}[hbtp]
  \centering
  \begin{equation*}
    \begin{array}{|c|c|c|c|c|c|c|}
      \hline
      \text{Spin} & \text{Energy} & \text{Charge} & \text{Mass}^2
      & \text{Name} & \text{Wave-f.} & \\
      \hline
      1 & E_0 + 1 & y & 4 E_0 (E_0-1) & W & f^{\lbrack 1; q;
        +\rbrack} & \\
      0 & E_0 + 2 & y & 4E_0 (E_0+1) & \Sigma & f^{\lbrack 0;
        q\rbrack} & \\
      0 & E_0 + 1 & y - 2 & 4E_0 (E_0 -1) & \pi & f^{\lbrack 3; q-4;
        +\rbrack} & \\
      0 & E_0 + 1 & y + 2 & 4E_0 (E_0-1) & \pi & f^{\lbrack 3; q+4;
        +\rbrack} & \\
      0 & E_0 + 1 & y & 4E_0 (E_0-1) & \pi & f^{\lbrack 3; q;
        +\rbrack} & \\
      0 & E_0 & y & 4(E_0-1)(E_0-2) & \phi & f^{\lbrack 2_s; q
        \rbrack}_* & p\\
      \hline
    \end{array}
  \end{equation*}
  \caption{The vector multiplet $W$. The normalization is $E_{\tgdb} + 4
    = E_0$, $y_{\tgdb} = y$.}
  \label{tab:vector_mul_long_W}
\end{table}

\begin{table}[hbt]
  \centering
  \begin{equation*}
    \begin{array}{|c|c|c|c|c|c|c|}
      \hline
      \text{Spin} & \text{Energy} & \text{Charge} & \text{Mass}^2
      & \text{Name} & \text{Wave-f.} &  \\
      \hline
      1 & E_0 + 1 & y & 4E_0 ( E_0 - 1) & Z & \chi^{\lbrack 2;
        q\rbrack} & \star, \square \\
      \frac{1}{2} & E_0 + \frac{1}{2} & y + 1 & -E_0 + 1 & \lambda_T
      & \chi_*^{\lbrack 1/2 \rbrack} & \star, \square \\
      0 & E_0 + 2 & y & 4 E_0 ( E_0 + 1) & \pi & \chi^{\lbrack 3; q;
        +\rbrack} & \\
      0 & E_0 + 1 & y - 2 & 4E_0 ( E_0 - 1) & \phi &
      \bar{\Omega}.\chi^{\lbrack 3; q; -\rbrack}_* & \star \\
      0 & E_0 + 1 & y + 2 & 4E_0 ( E_0 - 1) & \phi &
      \Omega.\chi^{\lbrack 3; q; +\rbrack}_* & \\
      0 & E_0 + 1 & y & 4E_0 ( E_0 - 1) & \phi & 
      J.\chi^{\lbrack 2; q\rbrack}_* & \star, \square \\
      0 & E_0 & y & 4(E_0-1) (E_0-2) & \pi & \chi^{\lbrack 3; q;
        -\rbrack} & \star, \square, p \\
      \hline
    \end{array}
  \end{equation*}
  \caption{The vector multiplet $Z$. ``$\star$'' marks modes remaining when
    $\chi$ is holomorphic. In this case $E_0 = y + 1$. The
    Betti multiplet is marked ``$\square$''. It corresponds to $d\chi = 0$
    and thus $y = 0$. The Lichnerowicz mode vanishes since there is no
    $(2,1)$ form to contract $\bar{\Omega}$ with. The normalization
    is $E_{\tgdb} + 1 = E_0$, $y_{\tgdb} = y$.}
  \label{tab:vector_mul_Z}
\end{table}

\begin{table}[hbtp]
  \centering
  \begin{equation*}
    \begin{array}{|c|c|c|c|c|c|c|}
      \hline
      \text{Spin} & \text{Energy} & \text{Charge} & \text{Mass}^2
      & \text{Name} & \text{Wave-f.} & \\
      \hline
      1 & E_0 + \frac{3}{2} & y - 1 & 4(E_0-\frac{1}{2})(E_0+\frac{1}{2}) & Z &
      \sigma^{\lbrack 2; q-4; -\rbrack} & \star \\
      1 & E_0 + \frac{3}{2} & y + 1 & 4(E_0-\frac{1}{2})(E_0+\frac{1}{2}) & Z &
      \sigma^{\lbrack 2; q; -\rbrack} & \\
      1 & E_0 + \frac{1}{2} & y - 1 & 4(E_0-\frac{3}{2})(E_0-\frac{1}{2}) & A &
      \sigma^{\lbrack 1; q-4\rbrack} & \star \\
      1 & E_0 + \frac{1}{2} & y + 1 & 4(E_0-\frac{3}{2})(E_0-\frac{1}{2}) & A &
      \sigma^{\lbrack 1; q\rbrack} & \star \\
      \frac{1}{2} & E_0 & y & E_0 - \frac{3}{2} & \lambda_L &
      \sigma^{\lbrack 1/2\rbrack}_* & \star, p \\
      0 & E_0 + \frac{3}{2} & y - 1 &
      4(E_0-\frac{1}{2})(E_0+\frac{1}{2}) & \phi & \sigma^{\lbrack
        2_s; q-4\rbrack}_*
      & \star^a \\
      0 & E_0 + \frac{3}{2} & y + 1 &
      4(E_0-\frac{1}{2})(E_0+\frac{1}{2}) & \phi & \sigma^{\lbrack
        2_s; q\rbrack}_*
      & \star^a \\
      0 & E_0 + \frac{1}{2} & y - 1 & 4(E_0-\frac{3}{2})(E_0-\frac{1}{2}) & \pi & 
      \sigma^{\lbrack 3; q-4; -\rbrack} & \\
      0 & E_0 + \frac{1}{2} & y + 1 & 4(E_0-\frac{3}{2})(E_0-\frac{1}{2}) & \pi &
      \sigma^{\lbrack 3; q; -\rbrack} & \\
      \hline
    \end{array}
  \end{equation*}
  \caption{The gravitino multiplet $\chi^+$ is constructed from
    $(1,0)$ forms. Holomorphy yields the short gravitino multiplet,
    the relevant modes are marked ``$\star$''. Once again, $E_{\tgdb} =
    y_{\tgdb}$. Out of the modes marked with ``$\star^a$'', one vanishes
    when $\sigma$ is holomorphic. $y_{\tgdb} = y + 1$,
    $E_{\tgdb} + \frac{1}{2} = E_0$. We included one of the fermionic
    modes, since it is the primary.}
  \label{tab:gravitino_mul_1}
\end{table}

\begin{table}[hbt]
  \centering
  \begin{equation*}
    \begin{array}{|c|c|c|c|c|c|c|}
      \hline
      \text{Spin} & \text{Energy} & \text{Charge} & \text{Mass}^2
      & \text{Name} & \text{Wave-f.} & \\
      \hline
      1 & E_0 + \frac{3}{2} & y - 1 & 4 (E_0+3)(E_0+2) & W &
      \sigma^{\lbrack 1; q-4\rbrack} & \\
      1 & E_0 + \frac{3}{2} & y + 1 & 4 (E_0+3)(E_0+2) & W &
      \sigma^{\lbrack 1; q\rbrack} & \\
      1 & E_0 + \frac{1}{2} & y - 1 & 4 (E_0+2)(E_0+1) & Z &
      \sigma^{\lbrack 2; q-4; +\rbrack} & \\
      1 & E_0 + \frac{1}{2} & y + 1 & 4 (E_0+2)(E_0+1) & Z &
      \sigma^{\lbrack 2; q; +\rbrack} & \\
      \frac{1}{2} & E_0 & y & - E_0 + \frac{3}{2} & \lambda_T &
      \sigma^{\lbrack 1/2 \rbrack}_* & p \\
      0 & E_0 + \frac{3}{2} & y - 1 & 4 (E_0+3)(E_0+2) & \pi &
      \sigma^{\lbrack 3; q-4; +\rbrack} & \\
      0 & E_0 + \frac{3}{2} & y + 1 & 4 (E_0+3)(E_0+2) & \pi &
      \sigma^{\lbrack 3; q; +\rbrack} & \\
      0 & E_0 + \frac{1}{2} & y - 1 & 4 (E_0+2)(E_0+1) & \phi &
      \sigma^{\lbrack 2_s; q-4\rbrack}_* & \\ 
      0 & E_0 + \frac{1}{2} & y + 1& 4 (E_0+2)(E_0+1) & \phi &
      \sigma^{\lbrack 2_s; q\rbrack}_* & \\
      \hline
    \end{array}
  \end{equation*}
  \caption{The gravitino multiplet $\chi^-$. $y_{\tgdb} = y + 1$,
    $E_{\tgdb} + \frac{5}{2} = E_0$. Again we include the fermionic primary.}
  \label{tab:gravitino_mul_2}
\end{table}

The modes derived from a scalar $f$ in table
\ref{tab:wave-functions_from_scalars} yield the graviton and vector
multiplets $A$ and $W$ in tables \ref{tab:graviton_multiplet},
\ref{tab:vector_mul_long_A}, and \ref{tab:vector_mul_long_W}. The
third family of vector multiplets, the vector multiplet $Z$
(table \ref{tab:vector_mul_Z}) is based on the primitive $(1,1)$ forms of
table \ref{tab:wave-functions_from_two-forms}. Modes based on one-forms (see
table \ref{tab:wave-functions_from_one-forms}) fill out the $\chi^+$
and $\chi^-$ gravitino multiplets (tables \ref{tab:gravitino_mul_1}
and \ref{tab:gravitino_mul_2}).  The appearance of $f$ in the short graviton multiplet and long vector $W$ is known as the ``shadow-mechanism'' \cite{Billo:2000zs}.

\paragraph{Short graviton, vector, and gravitino multiplets}
\label{sec:short-multiplets}

When $f$ is holomorphic, the inequality $E_{\tgdb} \geq y_{\tgdb}$ is
saturated. At the same time, several wave-functions vanish. Those
remaining are marked with a ``$\star$'' in table
\ref{tab:wave-functions_from_scalars}.
Comparing the table with tables \ref{tab:graviton_multiplet},
\ref{tab:vector_mul_long_A}, and \ref{tab:vector_mul_long_W}, we see
that the vector multiplet $W$ remains unchanged, while the graviton
multiplet shortens, becoming the short graviton multiplet. Out of the
modes forming the long vector multiplet $A$, only
\begin{equation}
  S\lbrack f^{\lbrack 0; q\rbrack} \rbrack \quad \text{and} \quad
  \pi\lbrack f^{\lbrack 3; q-4; -\rbrack} \rbrack
\end{equation}
remain. Again, $E_{\tgdb} = y_{\tgdb}$ and one sees that these modes form a
hypermultiplet as outlined in table \ref{tab:hyper_multiplet}.

The vector multiplet $Z$ shortens when $\chi$ is holomorphic. As shown
in appendix \ref{sec:lefschetz-decomposition}, all elements of
$H^{1,1}_\tgdb$ are primitive.

As mentioned in section \ref{sec:short-vector-and-betti-multiplets},
the holomorphic volume form $\Omega$ provides a map between scalars
and $(2,0)$ forms, which allows us to understand the shortening of the
vector multiplet $A$. Moreover, $\Omega$ allows us to map any
$(0,1)$ form $\alpha$ to a $(2,0)$ form $\alpha\lrcorner \Omega$. The
two-form is holomorphic if $\alpha$ is. However, $H^{0,1}_\tgdb = 0$
and thus any holomorphic $(2,0)$ form can be written as $\tgdb f
\lrcorner \Omega$. From our considerations in section
\ref{sec:holomorphy_discussion_for_chi} we know that holomorphic
$(1,1)$ forms saturate the inequality
\begin{equation}
    \Delta_2 \geq q_2(q_2+2),
\end{equation}
where we denote the charge of a two-from as $q_2$, to avoid
confusion with that of $f$, which we still refer to as $q$. Furthermore, we
know that the form $\tgdb f \lrcorner \Omega$ satisfies
\begin{equation}
  \Delta_2 (\tgdb f \lrcorner \Omega) = (\delta + 8) \tgdb f \lrcorner
  \Omega, \qquad
  \pounds_\xi (\tgdb f \lrcorner \Omega) = \imath (q+4) \tgdb f
  \lrcorner \Omega = \imath q_2 \tgdb f \lrcorner \Omega.
\end{equation}
Assuming that the bound holds for generic two-forms, we find
\begin{equation}
  \delta + 8 \geq (q+4)(q+6).  
\end{equation}
Solving for $\delta$ gives the shortening conditions for the
vector multiplet $A$:
\begin{equation}\label{eq:simple-shortening-conditino-vector-A}
  \tgdb (\tgdb f \lrcorner \Omega) = 0 \quad \text{and} \quad
  \delta = (q+2)(q+8) = 4 (y_{\tgdb} + 1)\left\lbrack (y_{\tgdb} + 1) + 3 \right\rbrack.
\end{equation}
Hence, even
though none of the modes in the long vector multiplet $A$ is directly
based on a two-form, the short vector multiplet $A$ corresponds to the
cohomology group $H^{2,0}_{\tgdb}$. As an aside, note that the
relation between the $\Delta_0$ and $\pounds_\xi$ eigenvalues in
\eqref{eq:simple-shortening-conditino-vector-A} is that of a
holomorphic function with charge $q+2$.

The $\chi^+$ and $\chi^-$ gravitino multiplets are shown in tables
\ref{tab:gravitino_mul_1} and \ref{tab:gravitino_mul_2}. When the
$(1,0)$ form $\sigma$ is holomorphic, various modes disappear as
discussed in \ref{sec:holomorphy_of_1,0-forms} and one obtains the
short gravitino multiplet as well as a long gravitino multiplet
$\chi^-$. We were not able to demonstrate the vanishing of the
three-form $\sigma^{\lbrack 3; q-4; -\rbrack}$ by direct manipulation
of the differential forms. According to the superconformal
algebra this mode should vanish.
An analysis of the complete multiplet including fermions and
Lichnerowicz modes should improve our understanding of the situation.
One can also construct the gravitino multiplets
in terms of $(0,1)$-forms $\tau$. As discussed in section
\ref{sec:wave-functions-derived-from-one-forms}, these cannot be
holomorphic, yet anti-holomorphic. When this happens, the $\chi^-$
multiplet shortens while the length of the $\chi^+$ multiplet remains
unaffected. Once again the mode $\tau^{\lbrack 3;  q+4; +\rbrack}$ remains.

\paragraph{The massless graviton multiplet}
\label{sec:massless-multiplets}

When $f$ is constant, it is necessarily also holomorphic and
antiholomorphic, so we can base the discussion on the holomorphic
case. Additional wave-functions vanish and the short graviton multiplet
shortens further to become the massless graviton multiplet.

\paragraph{Betti multiplets}
\label{sec:betti-multiplets}

Betti multiplets arise from a non-trivial second de Rham cohomology
group $H_{dR}^2(Y)$. Given $\alpha \in H_{dR}^2(Y)$, we know
that
\begin{equation}
  \tgdb \alpha = \tgd \alpha = \pounds_\xi \alpha = \Delta \alpha = 0.
\end{equation}
Assuming $\alpha \in \Omega^{2,0}$, the arguments of the previous
paragraphs imply that there is a scalar function $f$ such that $\alpha
= \tgdb f \lrcorner \Omega$ and that $\Delta \alpha = (\delta + 8)
\alpha$ with $\Delta f = \delta f$. However, $\delta \geq 0$ which is
in contradiction with $\alpha$ being harmonic. Hence, there are no
harmonic $(2,0)$ forms. An analogue argument excludes $(0,2)$
forms. Similarly, we can exclude forms in $\Omega^1 \wedge \eta$ since
$d(\sigma \wedge \eta) = d\sigma \wedge \eta - 2 \sigma \wedge J$ and
$\sigma \wedge J = 0$ has no solution. Turning to
$(1,1)$ forms, we can ignore non-primitive forms since
\begin{equation}
  0 = d(f J) = df \wedge J
\end{equation}
implies that $f$ has to be a constant. In the end, the only candidates
for Betti multiplets are the $(1,1)$ forms $\chi$ that are
included in the vector multiplet $Z$. In other words,
\begin{equation}\label{eq:deRham-Two_and_Kohn-Rossi11p}
  H_{dR}^2(Y) \subseteq H^{1,1}_{\tgdb}.
\end{equation}
For details on $H^{1,1}_\tgdb$ see appendix \ref{sec:lefschetz-decomposition}.

Finally, we can hazard a comment on Lichnerowicz modes in this
case. The construction of table \ref{tab:vector_mul_Z} assumes that
one way to obtain such modes is via symmetric contraction of $(2,1)$
forms with $\bar{\Omega}$. When $\tgd \chi = \tgdb \chi = 0$ 
the Lichnerowicz modes $\tgd\chi.\bar{\Omega}$ and $\tgdb\chi.\Omega$
vanish. Thus the multiplet has further shortening.

\paragraph{Hypermultiplets}
\label{sec:hypermultiplets}

\begin{table}[hbtp]
  \centering
  \begin{equation*}
    \begin{array}{|c|c|c|c|c|c|c|}
      \hline
      \text{Spin} & \text{Energy} & \text{Charge} & \text{Mass}^2
      & \text{Name} & \text{Wave-f.} & \\
      \hline
      0 & y + 1 & y - 2 & 4 y (y - 1) & \pi & f^{\lbrack 3;
        q-4; -\rbrack } & \\
      0 & y & y & 4 (y - 2 )(y-1) & S & f^{\lbrack 0;
        q\rbrack} & p \\
      \hline
      0 & y + 1 & - y + 2 & 4 y (y - 1) & \pi & \bar{f}^{\lbrack 3;
        q+4; +\rbrack} & \\
      0 & y & - y & 4 (y - 2)(y -1) & S & \bar{f}^{\lbrack 0;
        q\rbrack} & \\
      \hline
    \end{array}
  \end{equation*}
  \caption{The hypermultiplet. $f^{\lbrack 0; q\rbrack}$ and
    $f^{\lbrack 3; q-4; -\rbrack}$ are the ``survivors'' from the long
    vector multiplet $A$ when $f$ is holomorphic. The other two modes are
    their complex conjugates. As to the normalization, $E_0 = E_{\tgdb} = y_{\tgdb} = y$.}
  \label{tab:hyper_multiplet}
\end{table}

As we discussed in the previous paragraphs, most modes of the long
vector multiplet $A$ vanish when $f$ is holomorphic. The remaining modes form
a hypermultiplet as shown in table \ref{tab:hyper_multiplet}.

\begin{table}[hbtp]
  \centering
  \begin{equation*}
    \begin{array}{|c|c|c|c|c|c||c|c|c|}
      \hline
      \text{Spin} & \text{Energy} & \text{Charge} & \text{Mass}^2
      & \text{Name} & \text{Wave-f.} & \text{Name} & \text{Wave-f.} & \\
      \hline
      0 & y + 1 & y -2 & 4y (y-1) & \phi & \zeta^{\lbrack 2_s; q-4\rbrack}_*
      & \pi & \vartheta^{\lbrack 3; q \rbrack}  & \\
      0 & y & y & 4(y - 2)(y-1) & \pi & \zeta^{\lbrack 3; q\rbrack} & \phi & \vartheta^{\lbrack 2_s; q+4 \rbrack}_* & p\\
      \hline
      0 & y + 1 & - y + 2 & 4y (y-1) & \phi & \bar{\zeta}^{\lbrack
        2_s; q-4\rbrack}_* & \pi & \bar{\vartheta}^{\lbrack 3;
        q\rbrack}_* & \\
      0 & y & - y & 4(y - 2)(y-1) & \pi & \bar{\zeta}^{\lbrack 3;
        q\rbrack} & \phi & \bar{\vartheta}^{\lbrack 2_s;
        q+4\rbrack} & \\
      \hline
    \end{array}
  \end{equation*}
  \caption{Hypermultiplets constructed from $(2,1)$ forms $\zeta$ and
    $(1,2)$ forms $\vartheta$.}
  \label{tab:hyper_multiplet_2-1-form}
\end{table}

Since the hypermultiplet consists of two scalars, a spinor and their
complex conjugates, the primitive $(2,1)$ and $(1,2)$ forms of
section \ref{sec:extra-three-forms} also form hypermultiplets. Here,
the assumption that these forms can be mapped to 
eigenmodes of the Lichnerowicz operator via
\begin{equation}
  \zeta^{\lbrack 3; q\rbrack} \mapsto \zeta^{\lbrack 2_s; q-4\rbrack}
  = \zeta_{\kappa\lambda (\mu}
  \bar{\Omega}_{\nu)}^{\phantom{\nu)}\kappa\lambda}
  \qquad \text{and} \qquad
  \vartheta^{\lbrack 3; q\rbrack} \mapsto \vartheta^{\lbrack 2_s;
    q+4\rbrack} = \vartheta_{\kappa\lambda (\mu}
  \Omega_{\nu)}^{\phantom{\nu)}\kappa\lambda}
\end{equation}
is implied. The difference between the hypermultiplets in table
\ref{tab:hyper_multiplet_2-1-form} is the role reversal between the
Lichnerowicz and three-form modes. As a result the charge
of the $(1,2)$ multiplet is shifted when expressed in terms of the
cohomology group $H^{1,2}_\tgdb$.

\subsection{Previous Work on Kaluza-Klein Compactification}
\label{sec:previous-work}

The spectrum of Kaluza-Klein compactifications of eleven dimensional
supergravity to anti-de Sitter spaces is best understood for coset spaces where
harmonic analysis can be used. The multiplet tables in this paper are
based on those in \cite{Fabbri:1999mk}, thus our results agree with
theirs.

Going beyond harmonic analysis, \cite{Cassani:2012pj} and \cite{Gauntlett:2009zw}
consider consistent truncations of the
eleven-dimensional theory. It is instructive to compare our findings
with those of \cite{Cassani:2012pj}. Considering the reduction of
eleven-dimensional supergravity on seven-dimensional $SU(3)$-structure
manifolds, the 
authors of \cite{Cassani:2012pj} assume the existence of the following
real differentiable forms on $M_7$: A one-form $\theta$, $n_V$
two-forms $\omega_i$, $2n_H$ three-forms $\alpha_A$, $\beta^A$, $n_V$
four-forms $\tilde{\omega}^i$, and a six-form $\tilde{\omega}^0$. They
find that $n_V$ and $n_H$ correspond to the number of vector and hyper
multiplets in the four-dimensional theory. The $SU(3)$ structure is
expressed in terms of the above forms as
\begin{equation}
  \eta = e^V \theta, \quad
  J = e^{-V} v^i \omega_i, \quad
  \Omega = e^{-\frac{3}{2}V} (Z^A \alpha_A - \mathcal{G}_A \beta^A).
\end{equation}
Crucially, they impose a number of algebraic conditions on these
forms. First of all, the forms are all annihilated by the vector $k$ defined
by $\imath_k \theta = 1$ from which it follows that -- in the
Sasaki-Einstein case -- they are all elements of
$\Omega^{p,q}$. Furthermore, they require
\begin{equation}
  \omega_i \wedge \tilde{\omega}^j = -\delta_i^j \tilde{\omega}^0,
  \qquad
  \alpha_A \wedge \beta^B = -\delta_A^B \tilde{\omega}^0.
\end{equation}
Hence, one can think of $\tilde{\omega}^j \wedge \eta$ as the Hodge
dual of $\omega_i$. The same goes for $\beta^A \wedge \eta$ and
$\alpha_A$. Finally, the conditions
\begin{equation}
  \omega_i \wedge \alpha_A = \omega_i \wedge \beta^A =
  \tilde{\omega}^i \wedge \alpha_A = \tilde{\omega}^i \wedge \beta^A =
  \alpha_A \wedge \alpha_B = \beta^A \wedge \beta^B = 0,  
\end{equation}
are equivalent to our orthogonality conditions discussed in appendix
\ref{sec:details-of-supergravity-analysis}, since
\begin{equation}
  0 = \alpha_A \wedge \tilde{\omega}^i \wedge \eta \sim \alpha_A \wedge
  \star \omega_i
\end{equation}
it follows that all three-forms $\alpha_A$ and $\beta^A$ are
primitive. The same holds for $n_V -1$ two forms $\omega_i$, with the
only exception given by the linear combinatino that defines $J$. In
the Sasaki-Einstein case, the $2 n_H$ three-forms can be split into
$n_H - 1$ $(2,1)$ forms, one $(3,0)$ form and their complex
conjugates. Note that the forms $\omega_i$ cannot be $(2,0)$ or
$(0,2)$ since they vanish under the action of $\Omega \wedge$ and
$\bar{\Omega}\wedge$ and thus have to be of degree $(1,1)$.

Now, in our discussion we found one vector multiplet $Z$ for every
primitive $(1,1)$ form, which corresponds to $n_V - 1$ of the vector
multiplets. We also find a vector multiplet $W$ for a constant scalar
$f$, which corresponds to the non-primitive $(1,1)$ form $J$. This
gives a total of $n_V$ vector multiplets. Of course, we find additional vector
multiplets that do not appear in \cite{Cassani:2012pj} since we also
consider scalar fluctuations.

In section \ref{sec:hypermultiplets}, we found two sources of hyper
multiplets.  First there are holomorphic scalars, which are
equivalent to holomorphic $(3,0)$ forms. Second there are
holomorphic primitive $(2,1)$ forms.  We find more modes since our discussion includes scalar
fluctuations.

The field content of \cite{Gauntlett:2009zw} is that of a massless
graviton multiplet together with a long vector multiplet. In the
context of our paper, this corresponds to modes associated to constant
scalars $f$.

\section{Conclusions}
\label{sec:conclusions}

The main result of this paper is the computation of the gravity superconformal
index -- equation \eqref{aab}.
The supergravity calculation of the index is based on Kaluza-Klein
analysis of Sasaki-Einstein manifolds in section
\ref{sec:kaluza-klein-analysis}.  

A slight question remains regarding our analysis
of the short gravitino multiplet, where we were not able to show that
the three-form $\sigma^{\lbrack 3; q-4; -\rbrack}$ vanishes when
$\sigma$ is holomorphic and the multiplet shortens. This is a
shortcoming of our brute-force approach to constructing
wave-functions, since one has to argue their vanishing 
one-by-one. 
An analysis of the supergravity variations relating the modes
\cite{DAuria:1984vy} is a starting point towards a more satisfactory derivation.

There is a number of further problems in the supergravity sector that
we leave for future work. First, it is important to prove equation
\eqref{eq:deRham-KR-Laplacian-guess}.  Furthermore, there could possibly be additional shortening conditions.
Finally, it would be very interesting to determine the spectrum of short multiplets and the superconformal index of more general supergravity backgrounds than compactifications on Sasaki-Einstein manifolds.  This would give insight into theories whose holographic duals are even less well understood.

The superconformal index and the central charge $a$ of four-dimensional superconformal field theories are invariants of their associated Calabi-Yau 3-algebra \cite{ Eager:2012hx, Eager:2010yu}.  It would be very interesting to find a similar mathematical structure governing three-dimensional quiver Chern-Simons theories.  

The superconformal index has proven to be a powerful tool in checking proposed dualities.
While we have shown the equality of the field theory and gravity indices in several examples, much work remains to be done to show the equality with the field theory index for arbitrary geometries.  All proposed field theory duals to Saski-Einstein seven manifolds can be tested by computing the field theory index \cite{Imamura:2011su, Imamura:2011uj, Cheon:2011th} and comparing it with the gravity index in equation \eqref{aab}. 
Currently, there is no general procedure for constructing the field theory dual to a general Sasaki-Einstein seven manifold.  We hope that the superconformal index will help explore new holographic dualities.

\section*{Acknowledgements}
The authors would like to thank Jyotirmoy Bhattacharya, Alexey Bondal, Sergey Galkin, and Yuji Tachikawa for helpful discussions.
This work is  supported in part by World Premier International Research Center Initiative
(WPI Initiative),  MEXT, Japan through the Institute for the Physics and Mathematics
of the Universe, the University of Tokyo.

\appendix
\section{Conventions and Useful Expressions}\label{conventions}

We list our conventions for the calculations in sections
\ref{sec:multiplets} and appendix \ref{sec:supergravity}.

The Hodge star satisfies
\begin{equation}
  \begin{aligned}
    \star \alpha_{\mu_1 \dots \mu_{d-p}} &= \frac{\sqrt{g}}{p!}
    \epsilon_{\mu_1 \dots \mu_{d-p}}^{\phantom{\mu_1 \dots
        \mu_{d-p}}\nu_1 \dots \nu_p} \alpha_{\nu_1 \dots \nu_p}, \\
    \quad \star\star &= (-1)^{p(d-p)},
  \end{aligned}
\end{equation}
the generalization of the interior product $\lrcorner$
\begin{equation}
  \begin{aligned}
    \alpha_p \lrcorner \beta_q &= \frac{1}{p!} \alpha^{\mu_1 \dots \mu_p}
    \beta_{\mu_1 \dots \mu_p \nu_{q-p} \dots \nu_q} dx^{\nu_{q-p}}
    \otimes \dots \otimes dx^{\nu_q}, \\
    \star( \alpha \wedge \beta) &= \alpha \lrcorner \star \beta, \\
    \gamma_1 \lrcorner (\alpha_p \wedge \beta_q) &= (\gamma_1 \lrcorner
    \alpha_p) \wedge \beta_q + (-1)^p \alpha \wedge (\gamma_1
    \lrcorner \beta_q),
  \end{aligned}
\end{equation}
and the various kinetic operators are given by
\begin{equation}
  \begin{aligned}
    \Delta_0 \alpha &= - \nabla^\kappa \nabla_\kappa \alpha, \\
    \Delta_1 \alpha_\mu &= (- \nabla^\kappa \nabla_\kappa + 6) \alpha_\mu, \\
    \Delta_2 \alpha_{\mu\nu} &= - \nabla^\kappa \nabla_\kappa \alpha_{\mu\nu} -2
    R_{\kappa\mu\lambda\nu} \alpha^{\kappa\lambda} + 12
    \alpha_{\mu\nu}, \\
    Q\alpha_{\mu_1 \mu_2 \mu_3} &= (\star d \alpha)_{\mu_1 \mu_2
      \mu_3} = \frac{\sqrt{g}}{4!}
    \epsilon_{\mu_1\mu_2\mu_3}^{\phantom{\mu_1\mu_2\mu_3}\nu_1\dots\nu_4}
    d\alpha_{\nu_1\dots\nu_4}.
  \end{aligned}
\end{equation}
The operator $Q$ as defined here and $\boxtimes^{\lbrack 1, 1,
  1\rbrack}_y = M_{(1)^3}$ of \cite{DAuria:1984vy} are related as $Q =
4 \boxtimes^{\lbrack 1,1,1 \rbrack}_y$ since $\boxtimes_y^{\lbrack 1,1,1
  \rbrack} Y^{abc} = \frac{1}{4!} \epsilon^{abcd}_{\phantom{abcd}efg}
D_d Y^{efg}$.

\begin{table}[hbt]
  \centering
  \begin{equation*}
    \begin{array}{|cccc|}
      \hline
      \text{Scalars} & & & \\
      \pi & C_3 & Y_{\lbrack abc \rbrack} & Q^2 + 6 m Q + 8 m^2 \\
      \phi & h_{ab} & Y_{(ab)} & \Delta_L - 4 m^2 \\
      S & h^a_{\phantom{a}a}, C_3 & Y & \Delta_0 + 44m^2
      - 12 m \sqrt{ \Delta_0 + 9 m^2} \\
      \Sigma & h^a_{\phantom{a}a}, C_3 & Y & \Delta_0 + 44m^2
      + 12 m \sqrt{ \Delta_0 + 9 m^2} \\
      \hline
      \text{1-forms} & & & \\
      A & h_{ma} & C_{mna} & \Delta_1 + 12 m^2 - 6 \sqrt{\Delta_1
        + 4 m^2} \\
      W & h_{ma} & C_{mna} & \Delta_1 + 12 m^2 + 6 \sqrt{\Delta_1
        + 4 m^2} \\
      Z & C_3 & Y_{\lbrack a b \rbrack} & \Delta_2 \\
      \hline
      \text{Spin-2} & & & \\
      h & h_{\mu\nu}^{TT} & Y & \Delta_0 \\
      \hline
    \end{array}
  \end{equation*}
  \caption{The anti-de Sitter mass relations \cite{Duff:1986hr, Larsson2004}. The table lists the
    $AdS_4$ field, its 11-dimimensional origin, the resulting
    $7$-dimensional wave-function, and finally the mass operator. Here, $m$
    refers to the mass scale of the Freund-Rubin compactification, see
    \eqref{eq:Einstein-condition}, and is not to be confused with the
    mass of the bulk fields. In our conventions, $m = 1$. }
  \label{tab:mass-relations}
\end{table}

\begin{table}[htbp]
  \centering
  \begin{equation*}
    \begin{array}{|l|l|}
      \hline
      \text{Scalars} & \Delta_\pm = \frac{1}{2} ( 3 \pm \sqrt{9 + 4R^2
        (m^2 - 8)}) \\
      \text{p-Forms} & \Delta_\pm = \frac{1}{2} ( 3 \pm \sqrt{
          (3-2p)^2 + 4 R^2 m^2} ) \\
      \hline
    \end{array}
  \end{equation*}
  \caption{Scaling dimension and anti-de Sitter mass. Here, $m$ is the
    mass of the relevant bulk field. For the origin
    of the slightly awkward $-8$ term in the scalar expression, see
    footnote 5 in \cite{Gauntlett:2006vf}. Similarly, $R = 1/2$.}
  \label{tab:scaling_dim_from_ads_mass}
\end{table}

Turning to the properties of the Sasaki-Einstein links, we start with
the Einstein condition
\begin{equation}\label{eq:Einstein-condition}
  R_{\mu\nu} = 6m^2 g_{\mu\nu}.  
\end{equation}
Here, we chose $m = 1$. Moreover, the spaces inherit from the Calabi-Yau cone the
forms $\eta$, $J$, $\Omega$. The Reeb vector $\xi$ is given by
\begin{equation}
  \xi^\mu = g^{\mu\nu} \eta_\nu.
\end{equation}
The forms satisfy
\begin{equation}
  \begin{aligned}
    \nabla_\mu \eta_\nu &= J_{\mu\nu}, &
    \nabla_\kappa J_{\mu\nu} &= - \eta_\mu g_{\kappa\nu} + \eta_\nu
    g_{\kappa\mu}, &
    \nabla_{\kappa} \Omega_{\lambda\mu\nu} &= 4\imath
    \eta_{\lbrack\kappa} \Omega_{\lambda\mu\nu\rbrack},
  \end{aligned}    
\end{equation}
as well as
\begin{equation}
  \eta\lrcorner \eta = 1, \qquad
  \eta\lrcorner J = \eta\lrcorner\Omega = 0.
\end{equation}
It follows that $\eta$, $J$, $\Omega$ carry the charges $0$, $0$, $4$
under the Lie derivative along the Reeb vector $\pounds_\xi$. Using
the above relations, one can derive a number of useful contractions
involving the Riemann tensor, such as
\begin{equation}
  \begin{aligned}
    R_{\kappa\lambda\mu\nu} \eta^\nu &= g_{\kappa\mu} \eta_\lambda -
    g_{\lambda\mu} \eta_\kappa, \\
    J_\mu^{\phantom{\mu}o} R_{\kappa\lambda\nu o} -
    J_\nu^{\phantom{\nu}o} R_{\kappa\lambda\mu o} &= g_{\lambda\nu}
    J_{\kappa\mu} + g_{\kappa\mu} J_{\lambda\nu} - g_{\lambda\mu}
    J_{\kappa\nu} - g_{\kappa\nu} J_{\lambda\mu}, \\
    2 R_{\mu\kappa\nu\lambda} \Omega^{\mu\nu\tau} &=
    R_{\mu\nu\kappa\lambda} \Omega^{\mu\nu\tau}, \\
    R_{\mu\nu\kappa\lambda} \Omega^{\kappa\lambda\tau} &= 2
    \Omega_{\mu\nu}^{\phantom{\mu\nu}\tau}.
  \end{aligned}
\end{equation}
Finally,
\begin{equation}
  \begin{aligned}
    \star 1 = \frac{1}{3!} J^3 \wedge \eta, \quad
    \star \eta = \frac{1}{3!} J^3, \quad
    \star J = \frac{1}{2} \eta \wedge J^2, \quad
    \star\Omega = -\imath \Omega \wedge \eta.
  \end{aligned}
\end{equation}

As discussed in \cite{Eager:2012hx}, the cotangent space can be
decomposed as
\begin{equation}
  TY^* = T^{1,0}Y^* \oplus T^{0,1}Y^* \oplus \mathbb{C} \eta
\end{equation}
where $T^{1,0}Y^*$ is the eigenspace of $\Pi^+$:
\begin{equation}
  \begin{aligned}
    \Pi^\pm &= \frac{1}{2} ( g \mp \imath J - \eta \otimes \eta ).
  \end{aligned}
\end{equation}
Thus, generic $k$-forms can decomposed as ($k = p+q$)
\begin{equation}
  \Omega^k(Y) = \Omega^{p,q} \oplus (\Omega^{p,q-1} \oplus
  \Omega^{p-1,q}) \wedge \eta.
\end{equation}
Furthermore, the exterior derivative can be decomposed as
\begin{equation}
  d = \tgd + \tgdb + \eta\wedge \pounds_\xi.
\end{equation}
The tangential Cauchy-Riemann operators satisfy
\begin{equation}\label{eq:tCR-anti-commutator}
  \tgd \tgdb + \tgdb \tgd = -2 J \wedge \pounds_\xi.  
\end{equation}

\section{Details of the Supergravity Analysis}
\label{sec:details-of-supergravity-analysis}
\label{sec:supergravity}

In what follows, we will generally start with a eigenform $\alpha$ of the
Hodge-de Rham operator, and use it to construct further eigenforms of
$\Delta$ or $Q$. The procedure is quite straightforward;\footnote{
Some of the calculations get fairly involved. We found the Mathematica
package xAct extremely helpful.
\cite{Garcia:2007,MartinGarcia:2008qz}
}
one chooses a basis $v_i^{\lbrack \alpha; p; q\rbrack }$ and
diagonalizes the matrix
\begin{equation}
  \Delta v_i^{\lbrack \alpha; p; q\rbrack } = M_{ij}^{\lbrack \alpha; p; q\rbrack } v_j^{\lbrack \alpha; p; q\rbrack }.
\end{equation}
Here, $\alpha$ labels the mode we started with, $p$ the form degree and $q$
the $R$-charge of the new modes.

\subsection{Wave Functions constructed from Scalars}
\label{sec:modes-from-scalars}

Consider a scalar eigenmode of the Hodge-de Rham operator with
definite R-charge,
\begin{equation}
  \Delta f = \delta f, \qquad
  \pounds_\xi f = \imath q f = 2 \imath \hat{y}_0 f.
\end{equation}

At the level of one-forms, we consider the basis
\begin{equation}
  v^{\lbrack f; 1; q \rbrack}_i = \left\{ \eta f; \imath (\tgd - \tgdb) f; df \right\}
\end{equation}
and find
\begin{equation}
  M^{\lbrack f; 1; q \rbrack} =
  \begin{pmatrix}
    \delta + 12 & 2 & 0 \\
    2\delta & \delta & 2\imath q\\
    0 & 0 & \delta
  \end{pmatrix}.
\end{equation}
Diagonalization yields a gauge mode $df$ as well as the forms
$f^{\lbrack 1; q; +\rbrack}$ and $f^{\lbrack 1; q; -\rbrack}$.

Note that it is not possible to construct one-forms from $f$ with
charge $q\pm 4$ -- essentially, one would have to construct $(2,0)$ or
$(0,2)$ forms out of $f$ and contract them with $\Omega$ or its
conjugate. No such two-forms exist that are linear in $f$.

Proceeding to two-forms, we consider
\begin{equation}
  v_i^{\lbrack f; 2; q \rbrack} = \left\{ d\tgdb f, \eta \wedge df, f J, \eta \wedge
    (\tgd-\tgdb) f \right\},
\end{equation}
and find
\begin{equation}
  M^{\lbrack f; 2; q\rbrack} =
  \begin{pmatrix}
    \delta - 2q & \imath \lbrack q (q+6) - \delta \rbrack & - 2\imath
    \lbrack q (q+6) - \delta \rbrack & 0 \\
    4\imath & \delta + 2q + 8 & - 4q & 0 \\
    0 & - 2 & \delta + 12 & 0 \\
    0 & 2q & - 4q & \delta + 8.
  \end{pmatrix}
\end{equation}
Diagonalization gives two gauge modes -- $df^{\lbrack 1; q; \pm
  \rbrack}$ -- as well as $f^{\lbrack 2; q; a\rbrack}$ and $f^{\lbrack
  2; q; b\rbrack}$.

At the level of two-forms we find the first modes with shifted
charge. There is only one basis element in each case, so one can read off
$M^{\lbrack f; 2; q\pm 4 \rbrack}$ from
\begin{equation}
  \begin{aligned}
    \Delta (\tgdb f \lrcorner \Omega) &= (\delta + 8) \tgdb f
    \lrcorner \Omega, &
    \pounds_\xi (\tgdb f \lrcorner \Omega) &= \imath (q + 4) (\tgdb f
    \lrcorner \Omega), \\
    \Delta (\tgd f \lrcorner \bar{\Omega}) &= (\delta + 8) \tgd f
    \lrcorner \bar{\Omega}, &
    \pounds_\xi (\tgd f \lrcorner \bar{\Omega}) &= \imath (q - 4) (\tgd f
    \lrcorner \bar{\Omega}).
  \end{aligned}
\end{equation}

Turning to three-forms, we study $Q = \star d$ instead of
$\Delta$ with basis
\begin{equation}
  v_i^{\lbrack f; 3; q \rbrack} = \left\{ df \wedge J, f\eta \wedge J, (df \lrcorner J) \wedge
    J, \eta \wedge d(df \lrcorner J) \right\},  
\end{equation}
and
\begin{equation}
  M^{\lbrack f, 3, q\rbrack} =
  \begin{pmatrix}
    0 & 0 & 0 & 0 \\
    0 & 4 & -1 & 0 \\
    -\imath q & - \delta & 0 & -1\\
    -2\imath q & -2\delta & 0 & -2
  \end{pmatrix}.
\end{equation}
Again, there are two gauge modes and $f^{\lbrack 3; q; \pm\rbrack}$

Now, there are two different ways to construct a charged $(3,0)$ form:
$\tgd (\tgdb f \lrcorner \Omega)$ and $f\Omega$. It is reasonable to
assume that they are linearly related as long as $f$ is not
holomorphic. Contracting both with $\bar{\Omega}$, one finds that
\begin{equation}\label{eq:charged_three-forms_from_f_relation}
  \tgd (\tgdb f \lrcorner \Omega) = -\frac{\delta - q(q+6)}{2} f\Omega.
\end{equation}
We still include both modes in the basis,
\begin{equation}
  v_i^{\lbrack f; 3; q+4 \rbrack} = \left\{ f\Omega, \eta \wedge
    (\tgdb f\lrcorner \Omega), (\tgd - \tgdb) (\tgdb f \lrcorner
    \Omega), d(\tgdb f \lrcorner \Omega) \right\},
\end{equation}
and find
\begin{equation}
  M^{\lbrack f; 3; q+4\rbrack} =
  \begin{pmatrix}
    q+4 & \imath & 0 & 0 \\
    0 & 2 & \imath & 0 \\
    0 & -\imath (\delta + 8) & 0 & q+4 \\
    0 & 0 & 0 & 0
  \end{pmatrix}.
\end{equation}
There is one gauge mode and one additional mode with eigenvalue $q+4$
that is a remnant of the fact that our basis is not a basis. One also
finds two eigenmodes $f^{\lbrack 3; q+4; \pm\rbrack}$.

An identical calculation gives
\begin{equation}
  v_i^{\lbrack f; 3; q-4 \rbrack} = \left\{ f\bar{\Omega}, \eta \wedge
    (\tgd f\lrcorner \bar{\Omega}), (\tgd - \tgdb) (\tgd f \lrcorner
    \bar{\Omega}), d(\tgd f \lrcorner \bar{\Omega}) \right\},
\end{equation}
and
\begin{equation}
  M^{\lbrack f; 3; q-4\rbrack} =
  \begin{pmatrix}
    -(q-4) & - \imath & 0 & 0 \\
    0 & 2 & \imath & 0 \\
    0 & - \imath(\delta + 8) & 0 & q-4 \\
    0 & 0 & 0 & 0
  \end{pmatrix}.
\end{equation}
One can also deriv an equivalent relation for
\eqref{eq:charged_three-forms_from_f_relation} with the
anti-holomorphicity bound \eqref{eq:anti-holo-bound_f} appearing on
the right hand side.

\subsubsection{Shortening Conditions}
\label{sec:holomorphy-etc-for-scalars}

The spectrum for $f$ is actually bounded from below. Since
\begin{equation}\label{eq:bound_f_derived}
  \int \vol \bar{f} \Delta f = \int \vol \left\lbrack 2 \vert \tgdb f
    \vert^2 + q(q+6) \bar{f} f \right\rbrack,
\end{equation}
it follows that
\begin{equation}\label{eq:holo-bound_f}
  \delta \geq q(q+6)  
\end{equation}
with equality if and only if $f$ is holomorphic. Similarly, one finds that
antiholomorphic $f$ corresponds to the bound
\begin{equation}\label{eq:anti-holo-bound_f}
  \delta \geq q(q-6).  
\end{equation}
The latter is to be expected, since complex conjugation acts on the
charge as $q \mapsto - q$. Due to the Lichnerowicz obstruction
\cite{Gauntlett:2006vf}, holomorphic, non-constant $f$ satisfy $q \geq 1$.

We introduce $E_{\tgdb}$ via
\begin{equation}
  \delta = 4 E_{\tgdb} (E_{\tgdb} + 3),
\end{equation}
so \eqref{eq:holo-bound_f} amounts to $E_{\tgdb} \geq \frac{q}{2}$.

If $f$ is holomorphic the bound \eqref{eq:bound_f_derived} is
satisfied and many of the basis elements $v_i^{\lbrack f;
  p\rbrack}$ vanish. So do a number of wave functions:
\begin{equation}
  \label{eq:holo-f-goners}
  f^{\lbrack 1; q; -\rbrack}, \quad 
  f^{\lbrack 2; q; b\rbrack}, \quad
  f^{\lbrack 2; q+4\rbrack}, \quad
  f^{\lbrack 3; q; -\rbrack}, \quad
  f^{\lbrack 3; q+4; -\rbrack} \text{ or } f^{\lbrack 3; q+4; +\rbrack}.
\end{equation}
A number of remarks are in order here. First of all, neither of
$f^{\lbrack 2; q; a, b\rbrack}$ vanishes, yet they coincide. We simply label
the remaining mode $a$. Furthermore, out of the four
forms in $v^{\lbrack f; 3; q+4\rbrack}$, all except $f\Omega$
vanish. The latter is now an eigenform with eigenvalue $q+4$. Since
$\delta = q(q+6)$, this agrees with the eigenvalues of $f^{\lbrack
  3; q+4; +\rbrack}$ for $q \geq 0$. For $q < -6$ however, it agrees
with $f^{\lbrack 3; q+4; -\rbrack}$.

Anti-holomorphy of $f$ leads to the vanishing of
\begin{equation}
  \label{eq:anti-holo-f-goners}
  f^{\lbrack 1; q; +\rbrack}, \quad 
  f^{\lbrack 2; q; b\rbrack}, \quad
  f^{\lbrack 2; q-4\rbrack}, \quad
  f^{\lbrack 3; q; +\rbrack}, \quad
  f^{\lbrack 3; q-4; +\rbrack} \text{ or }
  f^{\lbrack 3; q-4; -\rbrack}.
\end{equation}
Again, $f^{\lbrack 2; q; a, b\rbrack}$ conincide while now all
elements of $f^{\lbrack f; 3; q-4\rbrack}$ except $f\bar{\Omega}$
vanish. Since this has eigenvalue $q-4$, it corresponds to the $-$
mode for $q \geq 6$ and to the $+$ mode for $q \leq 0$.

The spectrum simplifies further when $f$ is constant. Now, we have
$\delta = 0 = q$ while all modes except
\begin{equation}\label{eq:const-f-survivors}
  f^{\lbrack 0; 0 \rbrack}, \quad
  f^{\lbrack 1;0;+ \rbrack}, \quad
  f^{\lbrack 3;0;+ \rbrack}, \quad
  f^{\lbrack 3;+4;+ \rbrack}, \quad
  f^{\lbrack 3;-4;+ \rbrack}
\end{equation}
vanish. We ignore $f J = f^{\lbrack 2; 0; a\rbrack}$ which is also an
eigenmode, yet pure gauge.

Independent shortening conditions are given by
\begin{equation}
  \tgdb (\tgdb f \lrcorner\Omega) = 0,
\end{equation}
as well as $\tgd (\tgd f \lrcorner \bar{\Omega}) = 0$. See the
discussion following equation
\eqref{eq:simple-shortening-conditino-vector-A} for details.

\subsection{Wave Functions derived from One-forms}
\label{sec:wave-functions-derived-from-one-forms}

We next consider one forms that were not covered in the discussion in
section \ref{sec:modes-from-scalars}. They need to be orthogonal
to $\eta f$, $\tgd f$, $\tgdb f$. Moreover, such modes cannot be
mapped to scalars. In total one needs to impose\footnote{Note that the
primitivity condition on the exterior derivative -- $J \lrcorner
d\sigma = 0$ -- is equivalent to the vanishing of $\tgd^\dagger \sigma$ or
$\tgdb^\dagger \sigma$.}
\begin{equation}
  \eta \lrcorner \sigma = d^\dagger \sigma = \tgd^\dagger \sigma =
  \tgdb^\dagger \sigma = J \lrcorner d\sigma = 0.
\end{equation}
Finally, $\sigma$ must not be exact in terms of $d$, $\tgd$,
$\tgdb$. In what follows, we'll assume that
\begin{equation}
  \sigma^{\lbrack 1; q \rbrack} = \sigma \in \Omega^{1,0}, \qquad
  \Delta_1 \sigma = \delta \sigma, \qquad
  \pounds_\xi \sigma = \imath q \sigma = 2 \imath \hat{y}_0 \sigma.
\end{equation}

There is actually a second one form with identical Hodge-de Rham
eigenvalue $\delta$:
\begin{equation}
  \sigma^{\lbrack 1; q-4 \rbrack} = \tgd\sigma \lrcorner \bar{\Omega}.
\end{equation}

At the level of two-forms with charge $q$, we consider
\begin{equation}
  v_i^{\lbrack \sigma; 2; q \rbrack} = \left\{ \sigma \wedge \eta,
    \imath (\tgd - \tgdb) \sigma - \eta \wedge \sigma, d\sigma \right\}
\end{equation}
and find
\begin{equation}
  M^{\lbrack \sigma; 2; q \rbrack} =
  \begin{pmatrix}
    \delta + 10 & - 2 & 0 \\
    -2( \delta - 5) & (\delta - 2) & 2 \imath q \\
    0 & 0 & \delta
  \end{pmatrix}.
\end{equation}
The two non-gauge eigenmodes are listed in table
\ref{tab:wave-functions_from_one-forms}.

Two-forms with shifted charge $q-4$ are constructed from the $(0,1)$
form $\tgd\sigma \lrcorner \bar{\Omega}$:
\begin{equation}
  v_i^{\lbrack \sigma; 2; q-4\rbrack} = \left\{
    \eta\wedge (\tgd\sigma\lrcorner\bar{\Omega}),
    (\tgd-\tgdb)(\tgd\sigma\lrcorner\bar{\Omega}),
    d(\tgd\sigma\lrcorner\bar{\Omega}) \right\}.
\end{equation}
Then
\begin{equation}
  M^{\lbrack \sigma; 2; q-4\rbrack} =
  \begin{pmatrix}
    \delta + 8 & 2\imath & 0 \\
    -2\imath \delta & \delta & 2q-8 \\
    0 & 0 & \delta
  \end{pmatrix}.
\end{equation}
Once again, there are two modes with eigenvalues $\delta + 4 \pm 2
\sqrt{\delta + 4}$.

Three-forms of charge $q$ can be constructed from
\begin{equation}
  v_i^{\lbrack \sigma; 3; q \rbrack} = \left\{ J \wedge \sigma, \eta
    \wedge (\tgd-\tgdb) \sigma, d(\tgd-\tgdb- 2 \imath
    \eta \wedge ) \sigma \right\}.
\end{equation}
Diagonalizing
\begin{equation}
  M^{\lbrack \sigma; 3; q \rbrack} =
  \begin{pmatrix}
    q & 1 & 0 \\
    \delta -q^2 + 4 & -q & - \imath \\
    0 & 0 & 0
  \end{pmatrix},
\end{equation}
one finds the modes $\sigma^{\lbrack 3; q; \pm\rbrack}$ with
eigenvalues $\pm \sqrt{\delta + 4}$ in table
\ref{tab:wave-functions_from_one-forms}.

For three-forms with charge $q-4$, one uses the same construction
replacing $\sigma$ with $\tgd\sigma \lrcorner\bar{\Omega}$. Then
\begin{equation}
  v_i^{\lbrack \sigma; 3; q-4 \rbrack} = \left\{ J \wedge (\tgd\sigma
    \lrcorner\bar{\Omega}), \eta \wedge (\tgd-\tgdb) (\tgd\sigma
    \lrcorner\bar{\Omega}), d(\tgd-\tgdb - 2\imath \eta \wedge )
    (\tgd\sigma \lrcorner\bar{\Omega}) \right\}
\end{equation}
and
\begin{equation}
  M^{\lbrack\sigma; 3; q-4\rbrack} =
  \begin{pmatrix}
    -q + 4 & -1 & 0 \\
    -\delta + q(q-8) + 12 & q-4 & \imath \\
    0 & 0 & 0
  \end{pmatrix}.
\end{equation}
The eigenvalues are again $\pm \sqrt{\delta + 4}$.

\subsubsection{Shortening Conditions}
\label{sec:holomorphy_of_1,0-forms}

In principle, the Hodge-de Rham operator $\Delta_1$ should satisfy a
bound similar to \eqref{eq:holo-bound_f}. By contracting with $J$, one
can verify the equation
\begin{equation}
  \tgd\tgdb \sigma = \frac{\imath}{4} \lbrack \delta - q(q+4) \rbrack
  J \wedge \sigma,
\end{equation}
which suggests
\begin{equation}\label{eq:one-form-holo-bound}
  \delta \geq q(q+4).
\end{equation}
In light of \eqref{eq:one-form-holo-bound}, we define
$E_{\tgdb}$ via
\begin{equation} 
  \delta = 4 E_{\tgdb} (E_{\tgdb} + 2).
\end{equation}
The holomorphy bound is again given by $E_{\tgdb} \geq \frac{q}{2}$. Using
similar methods one can show that antiholomorhpic $(0,1)$ forms $\tau$ satisfy
\begin{equation}
  \tgdb\tgd \tau = - \frac{\imath}{4} \lbrack \delta - q(q-4) \rbrack
  J \wedge \tau,
\end{equation}
suggesting the bound
\begin{equation}\label{eq:one-form-anti-holo-bound}
  \delta \geq q(q-4).  
\end{equation}

If $\sigma$ is holomorphic, the basis elements in $v^{\lbrack \sigma;
  2, 3; q\rbrack}$ become linearly dependent and the modes
$\sigma^{\lbrack 2, 3; q; -\rbrack}$ vanish. Since $H^{1,0}_{\tgd} =
0$, none of the shortening conditions affects the $(0,1)$-form
$\tgd\sigma\lrcorner\bar{\Omega}$. If $\sigma$ is holomorphic, the
associated $(0,1)$ form $\tgd\sigma \lrcorner \bar{\Omega}$ is not anti-holomorphic\footnote{It
cannot be holomorphic since $H^{0,1}_{\tgdb} = 0$}. Since the fact
that $\sigma$ is holomorphic implies
\begin{equation}
  \Delta (\tgd\sigma\lrcorner \bar{\Omega}) = q(q+4)
  \tgd\sigma\lrcorner \bar{\Omega}, \qquad
  \pounds_\xi \tgd\sigma\lrcorner \bar{\Omega} = \imath(q-4)
  \tgd\sigma\lrcorner \bar{\Omega},
\end{equation}
yet \eqref{eq:one-form-anti-holo-bound} demands that
\begin{equation}
  \Delta (\tgd\sigma\lrcorner \bar{\Omega}) = (q-4) (q-8)
  \tgd\sigma\lrcorner \bar{\Omega},  
\end{equation}
which is clearly impossible.

\subsection{Wave Functions derived from Two-forms}
\label{sec:wave-functions-derived-from-two-forms}

Proceeding to higher from degree, we consider two-forms that were not
captured in the previous sections. See \cite{Pilch:2013fk} for a
similar, recent construction. Again we have to impose
orthogonality to previously constructed forms while also demanding that these forms
are not exact. Finally, it should not be possible to map them to forms
of lower degree. Hence we consider forms $\chi$ satisfying
\begin{equation}
  \chi \lrcorner \Omega = \chi \lrcorner \bar{\Omega} = \chi \lrcorner
  J = \eta \lrcorner \chi = d^\dagger \chi = \tgd^\dagger chi =
  \tgdb^\dagger \chi = 0.
\end{equation}
Note that $\chi$ is primitive.
\begin{equation}
    \chi^{\lbrack 2;q\rbrack} = \chi \in \Omega_p^{1,1}, \quad
    \Delta_2 \chi = \delta \chi, \quad
    \pounds_\xi \chi = \imath \chi.
\end{equation}

In order to construct three-forms, we use the basis
\begin{equation}
  v_i^{\lbrack \chi; 3; q\rbrack} = \left\{ \eta\wedge\chi,
    (\tgd-\tgdb) \chi, d\chi \right\}.
\end{equation}
Then the matrix
\begin{equation}
  M^{\lbrack \chi; 3; q\rbrack} =
  \begin{pmatrix}
    -2 & - \imath & 0 \\
    \imath \delta & 0 & -q \\
    0 & 0 & 0
  \end{pmatrix}
\end{equation}
yields two eigenmodes, listed in table
\ref{tab:wave-functions_from_one-forms}.

\subsubsection{Shortening Conditions}
\label{sec:holomorphy_discussion_for_chi}

Contracting four-forms with $J$, one finds
\begin{equation}
  J \lrcorner (J \wedge \chi) = \chi, \qquad
  J \lrcorner (d\tgdb\chi) = \imath \frac{\delta - q(q+2)}{2} \chi.
\end{equation}
Thus
\begin{equation}
  \tgdb\chi = 0 \quad \Rightarrow \quad
  \delta = q(q+2).
\end{equation}
Again we define $E_{\tgdb}$ accordingly,
\begin{equation}
  \delta = 4 E_{\tgdb} (E_{\tgdb} + 1) \quad \Rightarrow \quad
  E_{\tgdb} \geq \frac{q}{2}.
\end{equation}

Moreover, when $\chi$ is holomorphic, the basis elements in
$v_i^{\lbrack \chi; 3; q\rbrack}$ become linearly dependent. It turns
out that the mode $\chi^{\lbrack 3; q; +\rbrack}$ vanishes.

When $d\chi = 0$, $\chi$ is both holo- and anti-holomorphic with
vanishing charge $q$. Again $\chi^{\lbrack 3; q; +\rbrack}$ vanishes
and $q = \delta = 0$. As we argue in section
\ref{sec:betti-multiplets}, such forms are relevant for Betti
multiplets.

\subsection{Additional Three-form Modes}
\label{sec:extra-three-forms}

Finally, we consider the possibility of three-forms that have eluded
us. The same considerations as in sections
\ref{sec:wave-functions-derived-from-one-forms} and
\ref{sec:wave-functions-derived-from-two-forms} yield that such forms
are primitive, lie in $\Omega^{2,1} \oplus \Omega^{1,2}$.
They are closed\footnote{Otherwise $\zeta \in \Omega^{2,1}$ for example could
  be mapped to $\Omega^{1,1}\wedge \eta$ via $\star \tgdb\zeta$.}
under $\tgd$ and $\tgdb$, co-closed under $d$, $\tgd$, $\tgdb$ and
satisfy\footnote{Equation \eqref{eq:three-forms-Q-evaluation} holds
  without imposing this. However, it appears necessary to impose this
  rule in order to avoid overlap with the modes constructed in section
\ref{sec:wave-functions-derived-from-one-forms} since
\begin{equation*}
  (\zeta_{\kappa\lambda\mu}
  \bar{\Omega}_{\nu}^{\phantom{\nu}\kappa\lambda})
  \Omega^{\mu\nu}_{\phantom{\mu\nu}\rho} dx^\rho \in \Omega^{1,0}.
\end{equation*}
} $\zeta_{\kappa\lambda\lbrack\mu}
\bar{\Omega}_{\nu\rbrack}^{\phantom{\nu\rbrack}\kappa\lambda} = 0$.
Now, primitive, co-closed $(2,1)$ and $(1,2)$ forms are holomorphic if and only if
they are antiholomorphic. Since $\tgd\tgdb + \tgdb\tgd = -2 J \wedge
\pounds_\xi$, consistency requires that they either carry no charge or
are annihilated by the action of the Lefschetz operator $J
\wedge$. If they carry no charge, they are closed under the exterior
$d$. Assuming that
\begin{equation}\label{eq:three-forms-Q-evaluation}
  \zeta^{\lbrack 3; q\rbrack} = \zeta \in \Omega^{2,1}, \quad
  \vartheta^{\lbrack 3; q\rbrack} = \vartheta \in \Omega^{1,2}, \quad
  \pounds_\xi \zeta = \imath q \zeta,
  \pounds_\xi \vartheta = \imath q \vartheta,
\end{equation}
one finds that
\begin{equation}
  Q \zeta = \star d \zeta = - q \zeta \qquad \text{and} \qquad
  Q \vartheta = q \vartheta.
\end{equation}

\section{Comments on Lefschetz Decomposition and Kohn-Rossi
  Cohomology}
\label{sec:lefschetz-decomposition}

We remind the reader that on K\"ahler manifolds, Lefschetz
decomposition is the unique decomposition of $k$-forms in terms of
primitive forms $k-2h$ forms $a_{(h)}$:
\begin{equation}
  \alpha = \sum_{h = 0} a_{(h)} \wedge J^h.
\end{equation}
On K\"ahler manifolds, the decomposition is compatible with
cohomology.

Studying the standard proofs for the decomposition of forms
\cite{Voisin2008vol1}, \cite{Huybrechts2005}, it becomes clear that
the proof of decomposition also holds in the Sasaki-Einstein
case. I.e.~given a generic $k$-form $\alpha$, there are unique forms
$a^\perp_{(h)}$, $a^\parallel_{(h)}$ ($h = 0, 1, 2, \dots$) of degree $k-2h$ and $k-2h-1$ 
respectively and orthogonal to the Reeb vector $\xi$ such that
\begin{equation}
  \alpha = a^\perp_{(h)} \wedge J^h + \eta\wedge a^\parallel_{(h)} \wedge J^h. 
\end{equation}
This decomposition is not compatible with de Rham cohomology as
follows from the application of the exterior $d$:
\begin{equation}
  d\alpha = (da^\perp_{(h)} + 2 a^\parallel_{(h-1)}) \wedge J^h - \eta
  \wedge da^\parallel_{(h)} \wedge J^h.
\end{equation}
$d\alpha = 0$ requires that the $\alpha^\parallel_{(h)}$ are closed,
yet the same cannot be said for the $\alpha^\perp_{(h)}$.

Let us turn to Kohn-Rossi cohomology. Here, we only consider elements
of $\Omega^{p,q}_Y$ and thus all forms are annihilated by the action
of $\eta \lrcorner$. Hence we can drop the $\eta \wedge$ terms in the
decomposition (and thus also the $\perp$ subscripts). Acting with $\tgdb$,
\begin{equation}
  \tgdb\alpha = \sum_{h = 0}^{\lfloor k/2 \rfloor} \tgdb a_{(h)} \wedge J^h,
\end{equation}
and we find that $\alpha$ is $\tgdb$-closed if and only if the $a_{(h)}$
are. In what follows we will assume that this is the case (i.e.~that
$\alpha$ is $\tgdb$-closed). Noting that $\alpha$ is a $\pounds_\xi$
eigenmode if and only if the $a_{(h)}$ are, we assume also that
$\pounds_\xi a_{(h)} = \imath q a_{(h)}$ with $q \neq 0$. Finally, we
will assume that $a_{(0)} = 0$. Then, using
\eqref{eq:tCR-anti-commutator}, we find that $\alpha$ is $\tgdb$-exact.
\begin{equation}
  \tgdb \left(\frac{\imath}{2} q^{-1} \sum_{h = 1}^{\lfloor k/2
      \rfloor} \tgd a_{(h)} \wedge J^{h-1}\right)
  = \sum_{h = 1}^{\lfloor k/2 \rfloor} a_{(h)} \wedge J^h
  = \alpha.
\end{equation}
Thus we find the following result: \emph{All Kohn-Rossi cohomology
  classes $\lbrack \alpha\rbrack$ are either primitive or carry zero
  charge under $\pounds_\xi$.}
This is a somewhat typical result for Sasaki-Einstein geometry. The
$U(1)$-charge is an obstruction for the Lefschetz decomposition to
behave as on K\"ahler manifolds. It seems reasonable to expect that for forms with zero
charge Lefschetz decomposition extends to cohomology.
With this we conjecture that Kohn-Rossi cohomology groups allow for the
following decomposition, which makes use of the fact that the
$\pounds_\xi$ operator commutes with $\tgdb$, $J\wedge$ and their adjoints:
\begin{equation}
  H_\tgdb^{p,q} = \oplus_{\hat{q}\neq 0} \left\lbrack
  H_\tgdb^{p,q}\right\rbrack_{\text{primitive}}^{\pounds_\xi = i \hat{q}}
  \oplus_k \left\lbrack H_\tgdb^{p-k, q-k}\right\rbrack^{\pounds_\xi = 0}_{\text{primitive}}.
\end{equation}

In the case of $(1,1)$ forms, it follows immediately that all forms are
primitive, since the restriction of $H^{0,0}_\tgdb$ to elements
with zero charge is trivial:
\begin{equation}
  \label{eq:11-forms-are-primitive}
  H^{1,1}_\tgdb = \left\lbrack H^{1,1}_\tgdb
  \right\rbrack_{\text{primitive}} \oplus \left\lbrack
    H^{0,0}_\tgdb \right\rbrack^{\pounds_\xi = 0} = \left\lbrack H^{1,1}_\tgdb
  \right\rbrack_{\text{primitive}}.
\end{equation}

A similar result holds for $(2,1)$ forms. Here, the question is whether
holmorphic $(1,0)$ modes $\sigma$ with charge $0$ contribute to
$H^{2,1}_\tgdb$ via $\sigma \wedge J$. Since $\sigma$ is holomorphic,
the bound on the Laplace operator is satisfied. Since the charge is
zero, $\sigma$ is harmonic and thus closed. Hence, $\tgd\sigma =
0$. Since $H_\tgd^{1,0} = 0$, there is a scalar $f$ such that $\sigma =
\tgd f$. Thus, all elements of $H^{2,1}_\tgdb$ are primitive.

Interestingly, we can use the above construction to
\emph{locally} construct $(1,0)$ forms $j_f$ that satisfy $\tgdb j_f =
J$. Pick any scalar function $f$ that is holomorphic with respect to
$\tgdb$ and carries charge $q$. Then define
\begin{equation}
  j_f = \frac{\imath}{2} \tgd \log f^{q^{-1}}.
\end{equation}
Again, application of \eqref{eq:tCR-anti-commutator} gives the desired
result $\tgdb j_f = J$ locally.

\section{Cohomology using Borel-Weil-Bott}
\label{cohomology-appendix}
The twisted cohomology groups of homogenous spaces can be computed by an extension of the Borel-Weil-Bott theorem \cite{MR863714} \cite{Dionisi04}.  We summarize the results for complex projective space and the quadric $Q \in \CP^4.$
For $\CP^n$ with ample line bundle $\mathcal{L} = \cO(1),$ the ordinary cohomology groups are
$$
H^p(\CP^n, \Omega^q) =
\begin{cases}
\mathbb{C} & \text{if } $p = 0$\\
0 & \text{otherwise.}
\end{cases}
$$
For $\ell > 0,$ the twisted cohomology groups are
$$
H^p(\CP^3, \Omega^q(\ell)) =
\begin{cases}
\chi^{A_3}_{[\ell,0,0]} & \text{if } $(p,q) = (0,0)$\\
\chi^{A_3}_{[\ell-2,1,0]} & \text{if } $(p,q) = (0,1)$\\
\chi^{A_3}_{[\ell-4,0,1]} & \text{if } $(p,q) = (0,2)$\\
\chi^{A_3}_{[\ell-4,0,0]} & \text{if } $(p,q) = (0,3)$\\
0 & \text{otherwise.}
\end{cases}
$$
The quadric $Q \in \CP^4$ is equipped with the line bundle $\mathcal{L} = \cO_Q(1),$ which is the pullback of $\cO_{ \CP^4}(1)$ from the ambient projective space.
Its cohomology groups are
$$
H^p(Q, \Omega^q) =
\begin{cases}
\mathbb{C} & \text{if } $p = 0$\\
0 & \text{otherwise.}
\end{cases}
$$
For $\ell > 0,$ the twisted cohomology groups are
$$
H^p(Q, \Omega^q(\ell)) =
\begin{cases}
\chi^{SO(5)}_{[\ell,0]} & \text{if } $(p,q) = (0,0)$\\
\chi^{SO(5)}_{[\ell-2,2]} & \text{if } $(p,q) = (0,1)$\\
\chi^{SO(5)}_{[\ell-3,2]} & \text{if } $(p,q) = (0,2)$\\
\chi^{SO(5)}_{[\ell-3,0]} & \text{if } $(p,q) = (0,3)$\\
\bC & \text{if $(p,q) = (1,2)$ and $\ell = 1.$} \\
0 & \text{otherwise.}
\end{cases}
$$

\bibliographystyle{ytphys}
\small\baselineskip=.97\baselineskip
\bibliography{ref}

\end{document}